\documentclass[12pt,aps,preprint,prd,nofootinbib,groupedaddress]{revtex4-1}
\usepackage{hyperref}
\usepackage{amssymb}
\usepackage{amsmath}
\usepackage{bm}
\usepackage{graphicx}

\begin{document}

\title{Primordial Magnetic Helicity from Stochastic Electric Currents}
\author{Esteban Calzetta}
\email[E-mail me at: ]{calzetta@df.uba.ar}
\affiliation{Departamento de F\'\i sica and IFIBA, FCEyN - UBA, 
Ciudad Universitaria, CABA, Argentina.}
\author{Alejandra Kandus}
\email[E-mail me at: ]{kandus@uesc.br}
\affiliation{LATO - DCET - UESC - Rodovia Ilh\'{e}us-Itabuna km 16 s/n,\\
 Ilh\'{e}us - BA, Brazil}

\begin{abstract}
We study the possibility that primordial magnetic fields generated in the transition between inflation and reheating posses 
magnetic helicity, $H_M$. The fields are induced by stochastic currents of scalar charged particles created during the 
mentioned transition. We estimate the rms value of the induced magnetic helicity by computing different four-point SQED 
Feynman diagrams. For any considered volume, the magnetic flux across its boundaries is in principle non null, which means 
that the magnetic helicity in those regions is gauge dependent. We use the prescription given by 
Berger and Field and interpret our result as the difference between two magnetic configurations that coincide in the exterior 
volume. In this case the magnetic helicity gives only the number of magnetic links inside the considered volume. We calculate
a concrete value of $H_M$ for large scales and analyze the distribution of magnetic defects as a function of the scale. Those 
defects correspond to regular as well as random fields in the considered volume. We find that the fractal dimension of the distribution 
of topological defects is $D = 1/2$.  We also study if the regular fields induced on large scales are helical, finding 
that they are and that the associated number of magnetic defects is independent of the scale. In this case the fractal dimension
is $D=0$. We finally estimate the intensity of
fields induced at the horizon scale of reheating, and evolve them until the decoupling of matter and radiation under the hypothesis 
of inverse cascade of magnetic helicity. The resulting intensity is high enough and the coherence length long enough to have an 
impact on the subsequent process of structure formation.
\end{abstract}


\maketitle

\section{\label{intro}Introduction}

Large scale magnetic fields are widespread in the Universe. From galaxies to clusters of galaxies coherent magnetic 
fields are detected, with intensities that range from $\mu $Gauss to tenth of $\mu $Gauss. Our galaxy as well as nearby 
galaxies show magnetic fields coherent on the scale of the whole structure, while in galaxy clusters the coherent length 
is much less than the cluster's size \cite{carr-tay,bagchi-09}. A remarkable fact recently discovered by observations, 
is that high redshift galaxies also posses coherent fields with the same intensitis as present day galaxies 
\cite{high-z,wolfe-08,kronberg-apj-08}. This result challenges the generally accepted mechanism of magnetogenesis, namely 
the amplification of a primordial field of $\mathcal{O}\sim 10^{-31}-10^{-21}$ Gauss by a mean field dynamo 
\cite{moffatt,zeldovich,brand-02,bran-sub-05} acting during a time of the order of the age of the structure: either the 
primordial fields are more intense so the galactic dynamo saturates in a shorter time, or the dynamo does not work as it 
is currently thought. It is hoped that future observations of high redshift environments will shed more light on the features 
of primordial magnetic fields \cite{ska,lofar,edges}.

In view of the lack of success in finding a primordial mechanism for magnetogenesis that produces a sufficiently intense field, 
either to feed an amplifying mechanism, or to directly explain the observations (see Refs. \cite{kandus-11,ryu-12} 
as recent reviews), 
researchers began to delve on magnetohydrodynamical effects that could compensate the tremendous dilution of the field due to flux 
conservation during the expansion of the universe. Among the possibilities there is primordial turbulence 
\cite{son-99,gra-cal-02,cal-kan-10,giov-11}. 
Possible scenarios for it are the reheating epoch, the phase transitions (at least the electroweak one) and possibly the epoch of 
reionization, all dominated by out of equilibrium processes. 

A key ingredient to produce stable, large scale magnetic fields in three-dimensional MHD turbulence, is the transfer of magnetic 
helicity from small scales to large scales, at constant flux \cite{frish-75,pouquet-76} (see also Ref. 
\cite{mala-mull-13} and references therein). Magnetic helicity, $H_{M}$, is defined as the volume integral of the scalar product of 
the magnetic field $\mathbf{B}$ with the vector potential $\mathbf{A}$ \cite{berger,biskamp-03}. In three dimensions, and in the absence of ohmic 
dissipation, it is a conserved quantity that accounts for the non-trivial topological properties of the magnetic field \cite{berger}, 
such as the twists and links of the field lines. Unlike the energy that performs a natural, direct cascade, i.e., from large scales 
toward small ones where it is dissipated, magnetic helicity has the remarkable property of \emph{inverse cascading}, that is, magnetic 
helicity stored in small scales evolves toward larger scales \cite{frish-75,pouquet-76}. The fact that magnetic energy and magnetic 
helicity spectra are dimensionally related as $E_{k}^{M}\sim kH_{k}^{M}$ \cite{biskamp-03} produces a dragging of the former toward 
large scales, thus enabling the field to re-organize coherently at large scales \footnote{This mechanism however imposes severe constraints 
on the dynamo action. See Refs. \cite{bran-sub-05,blackman}}. 

It must be stressed that in a cosmological context, the inverse cascade mentioned above operates on scales of the order of the particle horizon 
or smaller. This is due to the fact that turbulence is a causal phenomenon. Magnetic helicity on the other hand can be 
induced at any scale, the topology of the fields then remains frozen if the scales are super-horizon and if there is no 
resistive decay. For subhorizon scales it is a sufficient condition for its conservation that the conductivity of the plasma 
be infinite \cite{biskamp-03}.

The interpretation of $H_{M}$ as the number of twists and links must be considered with care because from its
very definition it is clear that $H_{M}$ is gauge dependent. In their seminal work, Berger and Field 
\cite{berger} proved that if the field lines do not cross the boundaries of the volume of integration, i.e., the field 
lines close inside the considered volume, then $H_{M}$ as defined \emph{is} a gauge invariant quantity. These authors 
also addressed the case of open field lines, and wrote down a definition of gauge invariant magnetic helicity based on 
the difference of two such quantities for field configurations that have the same extension outside the considered volume. 
In this case the quantity obtained can be interpreted as the numbers of links inside the volume. 
In general it is not difficult to find Early Universe mechanisms that produce magnetic fields endowed with 
magnetic helicity: generation of helical magnetic fields has been already addressed in the framework of electroweak 
baryogenesis \cite{corn-97,vachas-01,copi-08,chu-11} and of leptogenesis \cite{long-13}. The main problem is still
in the low intensities obtained in more or less realistic scenarios.

The magnetic fields we consider in this work are induced by stochastic currents of scalar charges created gravitationally 
during the transition Inflation-Reheating \cite{ckm-98,kcmw-00,giov-shap-00} (see \cite{cal-hu-08} for more details), 
and such field configuration is of open lines. In the light of the analysis of Berger and Field, we shall discuss a criterion 
by which the result obtained can be considered as gauge invariant. 
The fields induced are random, the mean value of the magnetic helicity is zero, 
but not the corresponding rms deviation. We assume that those fields are weak enough to neglect their backreaction on the 
source currents, and show that the rms magnetic helicity can be written as the sum of four SQED Feynman graphs, one of them 
representing the mean value of $H_M$ and consequently identically null. The remaining three add to a non null value.
We compute the value of the helicity for large scales and find that the number density of links scales with the distance
$\kappa^{-1/2}$ from a given point as $\kappa^{5/2}$, which means that their fractal dimension is $D=1/2$ 
This number density takes into account defects due to both regular and random fields. We also calculate the value of $H_M$ due to
regular fields on a large scale. In this case the number density scales as $\kappa^3$, the corresponding fractal dimension being $D=0$. 
Using the relation $B^2\left(\kappa\right)\propto H_M\left(\kappa\right)\kappa$, we compare the
associated helical intensity to the one obtained by computing directly the correlation function of the magnetic field at
the same scale $\kappa^{-1}$. We find that both expressions coincide, which means that the fields generated by the
considered mechanism are indeed helical. We estimate the intensity of those smooth fields on a galactic scale, finding an
intensity too small to seed the dynamo. We finally address the
evolution of fields generated at scales of the order of the particle horizon at the end of reheating, through the
inverse cascade of magnetic helicity mechanism, until matter-radiation equilibrium. This evolution is based on the
assumption that during radiation dominance the plasma is in a (mild) turbulent state. We find that the number density of magnetic
links scales as $\kappa$, the corresponding fractal dimension then being $D=4$. The field intensity as well as the scale of coherence 
are in a range that could have and impact on the process of structure formation \cite{ryu-12}.

We work with signature $\left( -,+,+,+\right) $ and with natural units, i.e., $c=1=\hbar $, $e^2=1/137$. 
We use the Hubble constant during Inflation,  $H$, which we assume constant, to give dimensions to the different quantities, 
i.e. we consider spacetime coordinates $\left[ x\right] = H^{-1}$, Lagrangian density $\left[ \mathcal{L}\right] =H^{4}$, 
four vector potential $\left[ A^{\mu }\right] =H$, field tensor $\left[ F^{\mu \nu }\right] =H^{2}$, scalar field $\left[ \Phi \right]=H$.

The paper is organized as follows: Section \ref{sed} contains a brief description of scalar electrodynamics in curved spacetime. In Section
\ref{mh} we define magnetic helicity and describe briefly its main properties. In Section \ref{mhrf} we develope the formalism to study
magnetic helicity of random fields and estimate its rms value in different scenarios: In Subsection \ref{demh} we compute the SQED Feynman 
graphs that describe the magnetic helicity two-point correlation function. In Subsection \ref{dim} we provide some physical quantities relevant
for our study. In Subsection \ref{hm-tir} we describe the transition Inflation-Reheating and quote some useful formulae for our work.
In Subsection \ref{gd} we apply the analysis of Berger and Field to our fields and show the gauge invariance of our results. 
In Subsection \ref{mhls} we calculte the magnetic helicity rms value on large scales, and compute the density and fractal dimension 
of the distribution of defects. In Subsection \ref{hmgal} we compute the rms value of magnetic helicity due to solely smooth fields,
and find that the fields induced by the mechanism considered in this work are completely helical, but very weak. Finally in Subsection
\ref{hm-ss} we analyze the evolution of fields induced on scales of the order of the horizon at reheating along radiation dominance.
By considering conservation of magnetic helicity and assuming full inverse cascade is operative, we find at decoupling a magnetic field
of intensity and coherence that could impact on the process of structure formation. In Section \ref{dc} we sumarize and discuss our
results. We leave details of the calculations to the Appendices.

\section{\label{sed}Scalar electrodynamics in FRW}

In curved spacetime the action for a charged scalar field coupled to the electromagnetic field is given by 
\begin{equation}
S=-\int dt~d^{3}r~\mathcal{L}\left[ g^{\mu \nu },\Phi ,\Phi ^{\dag },A_{\mu } 
\right]  \label{a-1}
\end{equation}
with the Lagrangian density 
\begin{equation}
\mathcal{L}=\sqrt{-g}\left[ g^{\mu \nu }D_{\mu }\Phi D_{\nu }^{\dag }\Phi
^{\dag }+\left( m^{2}+\xi \mathcal{R}\right) \Phi \Phi ^{\dag }+\frac{1}{4}
F^{\mu \nu }F_{\mu \nu }\right]  \label{a-2}
\end{equation}
with $g^{\mu\nu}$ being the metric tensor that for a spatially flat Friedmann-Robertson-Walker spacetime reads $g^{\mu\nu}=diag\left( -1,a^2\left(t
\right),a^2\left(t\right),a^2\left(t\right)\right)$, $D_{\mu }=\partial _{\mu }-ieA_{\mu }$ the covariant derivative, $\mathcal{R}$ the scalar 
curvature, $\xi $ the coupling constant of the scalar field to the curvature and $F_{\mu \nu }=\partial _{\mu }A_{\nu }-
\partial _{\nu }A_{\mu }$ 
the electromagnetic field tensor. Due to the conformal invariance of the electromagnetic field in the spatially flat FRW universe, it is 
convenient to work with conformal time, defined as $d\tau =dt/a\left( t\right) $, with $t$ being the cosmological (or physical) time. 
The metric tensor then reads $g_{\mu \nu }=a^{2}\left( \eta \right) \eta _{\mu \nu }$, with $\eta _{\mu\nu }=diag\left( -1,1,1,1\right)$. 
We shall be dealing with fields in Inflation, Reheating and Radiation dominance. In those epochs the scale factors in physical time are 
respectively $a_{I}\left( t\right) =\exp \left( Ht\right)$, $a_{rh}\left( t\right) =\left( t/t_{0}\right) ^{2/3}$ and $a_{rad}\left( t\right) 
=\left( t/t_1\right)^{1/2}$, while in dimensionless conformal time, $\eta = H\tau$ they read $a_{I}\left( \eta \right) =\left( 1 -\eta\right)^{-1}$, 
$a_{rh}\left( \eta \right) =\left( 1+\eta /2\right) ^{2}$ and $a_{rad}\left(\eta\right) = \left( 1 + \eta\right)$. We note that $\eta = 0$ 
corresponds to the end of Inflation (in that epoch $\eta < 0$) and consequently the scale factor at that moment is $a_I\left(\eta = 0\right) =
a_{rh}\left(\eta = 0\right)= 1$. We rescale the fields according to
\begin{equation}
\Phi \rightarrow \frac{\varphi }{Ha\left( \eta \right) },\quad A_{\mu}\rightarrow \frac{\mathcal{A}_{\mu }}{H},\quad A^{\mu }\rightarrow 
\frac{\mathcal{A}^{\mu }}{Ha\left( \eta \right) },\quad B^{\mu }\rightarrow \frac{\mathcal{B}^{\mu }}{H^{2}a^{2}\left( \eta \right) },
\quad B_{\mu }\rightarrow \frac{\mathcal{B}_{\mu }}{H^{2}\left( \eta \right) }  \label{a-4}
\end{equation}
which means that $\varphi$, $\mathcal{A}_{\mu}$ and $\mathcal{B}_{\mu }$ are dimensionless. Working with the Coulomb gauge, 
$\partial _{i}\mathcal{A}^{i}=0$ (considering also $\mathcal{A}_{0}=0$), we obtain after taking variations of action (\ref{a-1}) the following 
evolution equations for the scaled scalar and electromagnetic fields
\begin{equation}
\partial _{\eta }^{2}\varphi -\nabla ^{2}\varphi +a^{2}\left( \eta \right)
\left( \frac{m^{2}}{H^{2}}+\xi \mathcal{R}-\frac{\ddot{a}\left( \eta \right) 
}{a^{3}\left( \eta \right) }\right) \varphi +e^{2}\mathcal{A}^{i}\mathcal{A}
_{i}\varphi -2ie\mathcal{A}^{i}\partial _{i}\varphi =0  \label{a-5}
\end{equation}%
\begin{equation}
\left( \partial _{\eta }^{2}-\nabla ^{2}\right) \mathcal{A}_{i}=\mathcal{J}
_{i}-2e^{2}\mathcal{A}_{i}\varphi \varphi ^{\dag }  \label{a-6}
\end{equation}
with $\mathcal{J}^{i}$ the electric current due to the scalar field, given by
\begin{equation}
\mathcal{J}^{i}=ie\eta ^{ij}\left[ \varphi \partial _{j}\varphi ^{\dag
}-\varphi ^{\dag }\partial _{j}\varphi \right]  \label{a-7}
\end{equation}
which, writing the complex field in term of real fields as $\varphi =
\left(\phi ^{1}+i\phi ^{2}\right) /\sqrt{2}$, can be expressed as
\begin{equation}
\mathcal{J}^{i}=e\eta ^{ij}\left[ \phi _{1}\partial _{j}\phi _{2}-\phi
_{2}\partial _{j}\phi _{1}\right]  \label{a-7b}
\end{equation}

In a first approximation we consider that the induced fields $\mathcal{A}^{i} $ are weak enough to discard their coupling to the scalar field 
given by the last two terms in eq. (\ref{a-5}) and the last term in eq. (\ref{a-6}). Also, we consider minimal coupling of scalar fields to 
gravity, for it will produce maximal particle creation \cite{ckm-98}. Eq. (\ref{a-5}) then turns into the Klein-Gordon equation for a free field 
in FRW universe,
\begin{equation}
\partial _{\eta }^{2}\varphi -\nabla ^{2}\varphi +\left[ a^{2}\left( \eta
\right) \frac{m^{2}}{H^{2}}-\frac{\ddot{a}\left( \eta \right) }{a\left( \eta
\right) }\right] \varphi =0  \label{a-5b}
\end{equation}
whose solutions are given in Appendix \ref{apa}.

For a realistic evolution of $\mathcal{A}^{i}$ dissipative effects must be taken into account. This we do by assuming that Ohm's law in 
its usual form, $\mathcal{J}^{i}=\sigma \left( \eta \right) \mathcal{E}^{i}$, is valid. In the lack of a clear knowledge about the early Universe 
plasma, we assume a traditional form for the electric conductivity considered in the literature, namely, that it is proportional to the plasma 
temperature, i.e., $\sigma \left( \eta \right) =\sigma_{0}T\left( \eta \right) H^{-2}e^{-2}$, with $T\left( \eta \right) =T_{0}/a\left( \eta \right)$
for a relativistic plasma. This assumption amounts to adding to the l.h.s. of eq.
(\ref{a-6}) a term of the form $\sigma \left( \eta \right) \partial _{\eta } \mathcal{A}^{i}$, and so the evolution equation for $\mathcal{A}_{i}$ 
reads
\begin{equation}
\left[ \frac{\partial ^{2}}{\partial \eta ^{2}}-\nabla ^{2}+\frac{\sigma _{0}}{H}\frac{\partial }{\partial \eta }\right] 
\mathcal{A}_{i}=\mathcal{J}_{i}\label{a-8}
\end{equation}
By taking curl of this equation we obtain the one corresponding to the
magnetic field, i.e.
\begin{equation}
\left[ \frac{\partial ^{2}}{\partial \eta ^{2}}-\nabla ^{2}+\frac{\sigma _{0}}{H}\frac{\partial }{\partial \eta }\right] 
\mathcal{B}_{i}=\epsilon_{ijk}\partial _{j}\mathcal{J}_{k}  \label{a-9}
\end{equation}
where $\epsilon _{ijk}$ is the Levi-Civita tensor density. Eqs. (\ref{a-8}) and (\ref{a-9}) can be readily integrated to give 
the fields in terms of their sources, i.e.
\begin{equation}
\mathcal{A}_{i}\left( \bar{x},\eta \right) =\int_{\eta _{0}}^{\eta }d\tau\int_{r_{0}}^{r}d^{3}s~G_{ret}\left( \bar{x},\bar{s},\eta ,\tau \right) 
\mathcal{J}_{i}\left( \bar{s},\tau \right)  \label{a-10}
\end{equation}
\begin{equation}
\mathcal{B}_{i}\left( \bar{x},\eta \right) =\int_{\eta _{0}}^{\eta }d\tau\int_{r_{0}}^{r}d^{3}s~G_{ret}\left( \bar{x},\bar{s},\eta ,\tau \right)
\epsilon _{ijk}\partial _{j}\mathcal{J}_{k}\left( \bar{s},\tau \right)
\label{a-11}
\end{equation}
with $G_{ret}\left( \bar{x},\bar{s},\eta ,\tau \right) $ the solution of (see Appendix \ref{apb})
\begin{equation}
\left[ \partial _{\eta }^{2}-\nabla ^{2}+\sigma _{0}\partial _{\eta }\right]G_{ret}\left( \bar{x},\bar{s},\eta ,\tau \right) 
=\delta \left( \bar{x}-\bar{s}\right) \delta \left( \eta -\tau \right)  \label{a-12}
\end{equation}
The electric currents we are considering consist of scalar charges created due to the change of the Universe's geometry during the 
transition Inflation-Reheating. This change makes the scalar field vacuum state during Inflation to correspond to a particle state in the 
subsequent phase \cite{birrel-94}. This process of ``particle creation'' is an out-of-equilibrium one, the resulting particle currents 
being stochastic. In the case of a charged field the mean value of the electric current is zero, but not its rms deviation, which sources
a random magnetic field, whose magnetic helicity we are going to compute.

\section{\label{mh}Magnetic Helicity}

Classically, magnetic helicity is defined as the volume integral of the scalar product between magnetic field and magnetic vector potencial 
\cite{moffatt,berger}, i.e.
\begin{equation}
\mathcal{H}_{M}\left( \eta \right) =\int_{V}d^{3}r~A^{i}\left( \bar{r},\eta\right) B_{i}\left( \bar{r},\eta \right)  \label{b-1}
\end{equation}
which in view of the dimensions of $A^{i}$ and $B^{i}$ it is already a dimensionlessl quantity. As stated in the Introduction, it is a measure 
of the non-trivial topology of the magnetic field inside the volume $V$, or in other words, eq. (\ref{b-1}) represents the number of twists 
and links of the field lines inside $V$. This interpretation, however, must be considered with care, because $\mathcal{H}_{M}$ is not gauge 
invariant unless the boundaries of the volume of integration are not intersected by $B$-lines, i.e. the field 
lines close inside $V$. On the other hand if the magnetic flux across the boundaries of $V$ is not 
zero, as in our case, it is still possible to define a gauge invariant measure of the links of the field inside $V$ \cite{berger} by considering 
the difference between two field configurations that coincide outside $V$. We shall discuss this issue in a following paragraph. Considering 
the conformal transformation of the fields and coordinates given by eq. (\ref{a-4}) and writing the physical volume as $V=a^{3}\left( \eta
\right) \mathcal{V}$, eq. (\ref{b-1}) can be written as 
\begin{equation}
\mathcal{H}_{M}\left( \mathcal{V},\eta \right) =\int_{\mathcal{V}}d^{3}x~\mathcal{A}^{i}\left( \bar{x},\eta \right) \mathcal{B}_{i}
\left( \bar{x},\eta \right)  \label{b-2}
\end{equation}
where we see that this is not diluted by the expansion, i.e., expansion does not erase or create the topology of the field, 
as it is to be expected. In classical magnetohydrodynamics $\mathcal{H}_{M}$ is one of the ideal invariants, i.e. it is a conserved 
quantity in the absense of ohmic dissipation \cite{frish-75,moffatt,biskamp-03}. This means that magnetic helicity cannot be created or 
destroyed by turbulence or non-dissipative evolution, being then a property of the magnetic field created at its birth. Unlike other 
ideally conserved quantities in 3-dimensional magnetohydrodynamics, magnetic helicity performs an \textit{inverse cascade} \cite{frish-75}. 
This means that instead of travelling towards small scales (large wavenumber) where it would be dissipated, it makes its way toward large 
scales (small wavenumbers), carrying with it a bit of magnetic energy. Mathematically, this is expressed by the spectral relation quoted
above, i.e., $B^2\left(\kappa\right)\propto H_M\left( k\right) k$. In a framework of decaying turbulence (that could exist at the earliest
epochs of the Universe), there would be a 
self-organization of the magnetic field at large scales, with the total energy contained in the considered volume decaying as
$E\propto E_0 \eta^{-2/3}$ and coherence length increasing as $\lambda\propto \lambda_0\eta^{2/3}$ \cite{biskamp-03}.
For primordial magnetogenesis this fact can be of great help in obtaining stronger fields than the ones found up to now to seed subsequent 
amplifying mechanisms, or even to directly explain the observations. But on the other hand, the conservation of $\mathcal{H}_M$ crucially 
constraints the operation of further amplifying mechanisms such as the mean field dynamo \cite{blackman}.

\section{\label{mhrf}Magnetic Helicity of Random Fields}

As we are dealing with random fields, generated from stochastic quantum electric currents whose mean value is zero, we must evaluate a rms 
value of the helicity by calculating a two-point correlation function given by
\begin{equation}
\Xi \left( \mathcal{V},\eta ,\eta ^{\prime }\right) \equiv \left\langle 
\mathcal{H}_{M}\left(\mathcal{V}, \eta \right) ,\mathcal{H}_{M}\left( \mathcal{V} 
\eta ^{\prime}\right) \right\rangle   \label{b-3}
\end{equation}
where angle brackets denote stochastic and quantum average. We shall consider the volume of integration $\mathcal{V}$ as the commoving space occupied by the structure of interest (i.e., a galaxy, a cluster, particle horizon at a certain epoch, etc.).

\subsection{\label{demh}Diagramatic evaluation of the magnetic helicity}

We begin by writing
\begin{equation}
\Xi\left( \mathcal{V},\eta ,\eta ^{\prime }\right)= \int d^{3}\kappa^{\prime}\int d^{3}\kappa \int_{\mathcal{V}}d^{3}x^{\prime}
\int_{\mathcal{V}}d^{3}x \exp{\left(i\bar{\kappa}\cdot \bar x\right)} 
\exp{\left(i\bar{\kappa}^{\prime}\cdot \bar x^{\prime}\right)} \Xi\left(\bar\kappa , \bar \kappa^{\prime},\eta , \eta^{\prime}\right)
\label{c-1}]
\end{equation}
where
\begin{equation}
\Xi\left(\bar\kappa , \bar \kappa^{\prime},\eta , \eta^{\prime}\right)=\mathcal{H}_M\left(\bar{\kappa},\eta\right) 
\mathcal{H}_M\left(\bar {\kappa}^{\prime},\eta^{\prime}\right)  \label{c-2}
\end{equation}
with 
\begin{equation}
\mathcal{H}_{M}\left( \bar{\kappa},\eta \right) =~\mathcal{A}_{i}\left( \bar{p},\eta \right) \mathcal{B}^{i}
\left( \bar{\kappa}-\bar{p},\eta \right)
\label{c-3}
\end{equation}
To avoid cumbersome notation, repeated momenta other than $\kappa$ are assumed to be integrated over.
The volume integrals can be readily evaluated, giving
\begin{eqnarray}
\int_{\mathcal{V}}d^{3}x \exp{\left(i\bar{\kappa}\cdot \bar x\right)} &\simeq &  \mathcal{V}\qquad \textrm{for }~~ \bar{\kappa}\cdot \bar x 
\lesssim 1 \label{c-4} \\
&\simeq & 0 \qquad \textrm{for }~~ \bar{\kappa}\cdot \bar x > 1\nonumber
\end{eqnarray}
The different fields in (\ref{c-3}) can be written as
\begin{equation}
\mathcal{A}_{i}\left( \bar{p},\eta \right) =G_{ret}\left( \eta -\tau_{1},\bar p\right) \delta \left(\bar p-\bar p_{1}\right) \mathcal{J}_{i}
\left( \bar p_{1},\tau_{1}\right)  \label{c-5}
\end{equation}
and
\begin{equation}
\mathcal{B}_{i}\left( \bar{\kappa}-\bar{p},\eta \right) =i\epsilon^{ils}G_{ret}\left( \eta -\tau _{2},\bar\kappa - \bar p\right) 
\delta \left(\bar \kappa-\bar p-\bar p_{2}\right) \left( \kappa _{l}-p_{l}\right) \mathcal{J}_{s}\left( \bar p_{2},\tau _{2}\right)  \label{c-6}
\end{equation}
(integration in $\tau ^{\prime }s$ is understood). The electric currents $\mathcal{J}_{i}\left(\bar  p,\tau \right) $ can be expressed in terms
of the scalar fields as 
\begin{equation}
\mathcal{J}_{i}\left( \bar p,\tau \right) =ie\delta \left( \bar q_{1}+\bar q_{2}-\bar p\right)\delta \left( \tau -\varsigma _{1}\right) 
\delta \left( \tau -\varsigma_{2}\right) \left( q_{1i}-q_{2i}\right) \phi ^{1}\left( \bar q_{1},\varsigma_{1}\right) \phi ^{2}
\left(\bar q_{2},\varsigma _{2}\right)  \label{c-7}
\end{equation}
Gathering all expressions, $\mathcal{H}_M\left(\bar{\kappa},\eta\right)$ is written as
\begin{eqnarray}
\mathcal{H}_{M}\left( \bar{\kappa},\eta \right) &=&-ie^{2}G_{ret}\left( \eta-\tau _{1},\bar p\right) G_{ret}\left( \eta -\tau _{2},
\bar \kappa -\bar p\right) 
\delta\left( \tau _{1}-\varsigma _{1}\right) \delta \left( \tau _{1}-\varsigma_{2}\right) \delta \left( \tau _{2}-\varsigma _{3}\right) 
\delta \left( \tau_{2}-\varsigma _{4}\right)  \notag \\
&&\delta \left( \bar p-\bar p_{1}\right) \delta \left(\bar  \kappa -\bar p-\bar p_{2}\right) \delta\left(\bar  q_{1}+\bar q_{2}-\bar p_{1}\right) \delta \left( \bar q_{3}+\bar q_{4}-\bar p_{2}\right)
\label{c-8} \\
&&\epsilon ^{ils}\left( q_{1i}-q_{2i}\right) \left( q_{3s}-q_{4s}\right)\left( \kappa _{l}-p_{l}\right) 
\phi ^{1}\left( \bar q_{1},\varsigma _{1}\right)
\phi ^{2}\left( \bar q_{2},\varsigma _{2}\right) \phi ^{1}\left( \bar q_{3},\varsigma_{3}\right) 
\phi ^{2}\left(\bar  q_{4},\varsigma _{4}\right)  \notag
\end{eqnarray}
After integrating out the time delta functions the magnetic helicity correlation function spectrum reads 
\begin{eqnarray}
\Xi \left( \bar{\kappa} ,\bar{\kappa}^{\prime } ,\eta ,\eta^{\prime }\right)
&=&-e^{4}G_{ret}\left( \eta -\tau _{1},\bar p\right) G_{ret}\left( \eta -\tau_{2},\bar{\kappa} -\bar p\right) 
G_{ret}\left( \eta ^{\prime }-\tau_{1}^{\prime },\bar p^{\prime }\right) 
G_{ret}\left(\eta^{\prime }-\tau_{2}^{\prime },\bar{\kappa}^{\prime } -\bar p^{\prime }\right)  \notag \\
&&\delta \left( \bar p-\bar p_{1}\right) \delta \left(\bar{\kappa} -\bar p -\bar p_{2}\right) 
\delta \left( \bar p^{\prime }-\bar p_{1}^{\prime }\right)\delta \left( \bar{\kappa}^{\prime } - \bar p^{\prime }-p_{2}^{\prime }\right)
\label{c-9} \\
&&\delta \left( \bar q_{1}+\bar q_{2}-\bar p_{1}\right) \delta \left(\bar q_{3}+\bar q_{4}-\bar p_{2}\right) 
\delta \left( \bar q_{1}^{\prime }+\bar q_{2}^{\prime}-\bar p_{1}^{\prime }\right) 
\delta \left( \bar q_{3}^{\prime }+\bar q_{4}^{\prime}-\bar p_{2}^{\prime }\right)  \notag \\
&&\epsilon ^{abc}\left( q_{1a}- q_{2a}\right) \left(  q_{3c}- q_{4c}\right)\left(  \kappa _{b}- p_{b}\right) 
\epsilon ^{def}\left( q_{1d}^{\prime}-q_{2d}^{\prime }\right) \left( q_{3f}^{\prime }-q_{4f}^{\prime }\right)
\left( \kappa^{\prime }_{e}-p_{e}^{\prime }\right)  \notag \\
&&\left\langle \phi ^{1}\left( \bar q_{1},\tau_{1}\right) \phi ^{1}\left(\bar q_{3},\tau_{2}\right) \phi ^{1}\left( \bar q_{1}^{\prime },
\tau _{1}^{\prime}\right) \phi ^{1}\left( \bar q_{3}^{\prime },\tau_{2}^{\prime }\right)\right\rangle \nonumber\\
&&\left\langle \phi ^{2}\left( \bar q_{2},\tau _{1}\right) \phi^{2}\left( \bar q_{4},\tau_{2}\right) \phi ^{2}\left( \bar q_{2}^{\prime},
\tau_{1}^{\prime }\right) \phi ^{2}\left( \bar q_{4}^{\prime },\tau_{2}^{\prime}\right) \right\rangle  \notag
\end{eqnarray}
Each scalar field can be decomposed in its positive and negative frequency components, that respectively include the annihilation and 
the creation operator, namely 
\begin{equation}
\phi ^{i}\left( \bar q,\tau \right) =\phi ^{+}\left(\bar q,\tau \right) +\phi^{-}\left( \bar q,\tau \right) =\phi \left( \bar q,\tau \right) a_{q}
+\phi ^{\ast}\left( \bar q,\tau \right) a_{-q}^{\dag }  \label{c-10}
\end{equation}
When replacing this decomposition in expression (\ref{c-9}) it can be seen that in each bracket the only terms that contribute to the mean value are
\begin{eqnarray}
&&\left\langle \phi ^{1}\left( \bar q_{1},\tau_{1}\right) \phi ^{1}\left(\bar q_{3},\tau_{2}\right) \phi ^{1}\left( \bar q_{1}^{\prime },
\tau _{1}^{\prime}\right) \phi ^{1}\left( \bar q_{3}^{\prime },\tau_{2}^{\prime }\right)
\right\rangle \rightarrow \notag\\
&&\left\langle \phi ^{+}\left(\bar q_{1},\tau_{1}\right) \phi^{+}\left(\bar q_{3},\tau_{2}\right) \phi ^{-}\left( \bar q_{1}^{\prime },
\tau_{1}^{\prime }\right) \phi ^{-}\left( \bar q_{3}^{\prime},\tau_{2}^{\prime }\right) \right\rangle  \label{c-11} \\
&&+\left\langle \phi ^{+}\left( \bar q_{1},\tau_{1}\right) \phi ^{-}\left(\bar q_{3},\tau_{2}\right) \phi ^{+}\left( \bar q_{1}^{\prime },
\tau_{1}^{\prime}\right) \phi ^{-}\left( \bar q_{3}^{\prime },\tau_{2}^{\prime }\right)
\right\rangle  \notag
\end{eqnarray}
which in the end combine to form the scalar positive frequency operator, 
\begin{equation}
G^{+}\left(\bar q,\bar q^{\prime },\tau ,\tau^{\prime }\right) =\phi\left(\bar q,\tau \right) 
\phi ^{\ast }\left(\bar q^{\prime },\tau^{\prime}\right) \delta \left( \bar q+\bar q^{\prime }\right)  \label{c-12}
\end{equation}
The first term on the r.h.s. of expression (\ref{c-11}) gives 
\begin{eqnarray}
&&\left\langle \phi ^{+}\left( \bar q_{1},\tau_{1}\right) \phi ^{+}\left(\bar q_{3},\tau_{2}\right) 
\phi ^{-}\left( \bar q_{1}^{\prime },\tau _{1}^{\prime}\right) 
\phi ^{-}\left( \bar q_{3}^{\prime },\tau_{2}^{\prime }\right)\right\rangle =\notag \\
&&\frac{1}{2}G^{+}\left( \bar q_{1},\bar q_{1}^{\prime},\tau_{1},\tau_{1}^{\prime }\right)
 G^{+}\left(\bar  q_{3},\bar q_{3}^{\prime},\tau_{2},\tau_{2}^{\prime }\right)  \label{c-13} \\
&&+\frac{1}{2}G^{+}\left( \bar q_{1},\bar q_{3}^{\prime },\tau_{1},\tau _{2}^{\prime}\right) 
G^{+}\left( \bar q_{3},\bar q_{1}^{\prime },\tau _{2},\tau_{1}^{\prime}\right)  \notag
\end{eqnarray}
while the second term gives 
\begin{equation}
\left\langle \phi^{+}\left( \bar q_{1},\tau_{1}\right) \phi^{-} \left(\bar q_{3},\tau_{2}\right)
 \phi^{+}\left( \bar q_{1}^{\prime },\tau _{1}^{\prime}\right) 
\phi^{-}\left( \bar q_{3}^{\prime },\tau_{2}^{\prime} \right)\right\rangle =
G^{+}\left( \bar q_{1},\bar q_{3},\tau_{1},\tau_{2}\right) 
G^{+}\left(\bar q_{1}^{\prime },\bar q_{3}^{\prime },\tau_{1}^{\prime },\tau_{2}^{\prime }\right)
\label{c-14}
\end{equation}
When performing the products of the two brackets we obtain nine terms that
can be represented by the following graphs (full lines indicate scalar
fields, dotted lines vector potential and dashed lines magnetic fields).

\begin{center}
\begin{figure}[h]
\includegraphics[width=2.5in]{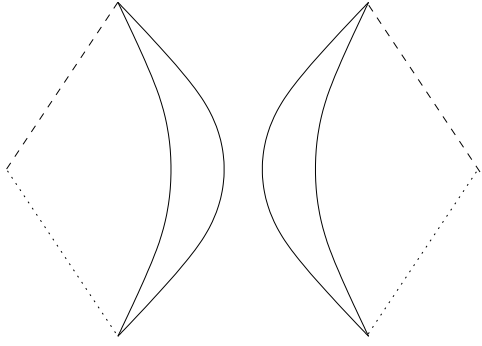}
\caption{``Mean Helicity'' graph.}
\end{figure}

\begin{figure}[h]
\includegraphics[width=2.5in]{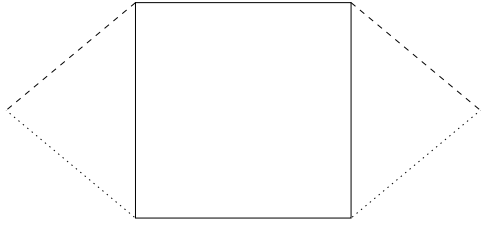}
\caption{``Square'' graph, $\Xi_{s}$}
\end{figure}

\begin{figure}[h]
\includegraphics[width=2.5in]{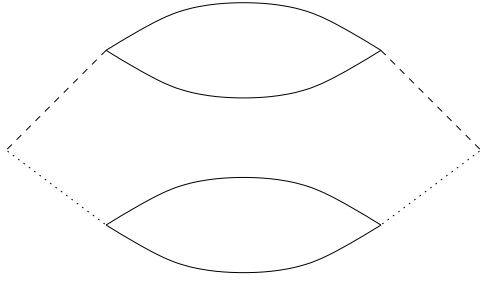}
\caption{``Two Bubbles'' graph, $\Xi_{2b}$}
\end{figure}

\begin{figure}[h]
\includegraphics[width=2.5in]{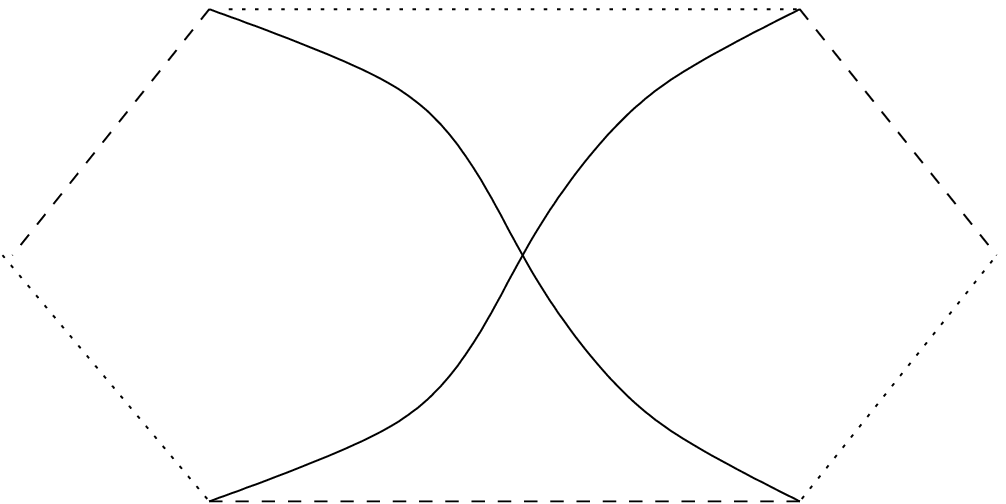}
\caption{``Cross'' graph, $\Xi_{c}$\footnote{Scalar field lines do not intersect in the middle of the graph.}}
\end{figure}
\end{center}

Observe that of the two vertices of scalar electrodynamics, only one contributes to these graphs: the one similar to the QED vertex. 
This is so because we disregarded the backreaction of $A^i$ on the scalar fields. Observe also that the first graph, being the product 
of two mean values, vanishes identically. The multiplicity of the ``square'' graph is 4, of the ``cross'' diagram is 2 and
of the ``two-bubble'' figure is also 2.

After replacing the expressions for the scalar positive frequency operators and solving all the Dirac delta functions we obtain a 
$\delta\left(\bar\kappa + \bar\kappa^{\prime }\right)$, which means that the momentum of the electromagnetic field is conserved.
Writing each graph as 
\begin{equation}
\Xi\left(\bar\kappa , \bar\kappa^{\prime},\eta ,\eta^{\prime}\right) = \xi\left(\bar\kappa ,\eta ,\eta^{\prime}\right)
\delta\left(\bar\kappa + \bar\kappa^{\prime}\right)\label{c-150}
\end{equation} 
we have the following expressions for the prefactor of the non-null graphs: 
\begin{eqnarray}
\xi_{s}\left(\bar \kappa ,\eta ,\eta ^{\prime }\right)&=&
-2e^{4}G_{ret}\left( \eta -\tau _{1},\bar{p}\right) G_{ret}\left( \eta
-\tau _{2},\bar{\kappa}-\bar{p}\right) G_{ret}\left( \eta ^{\prime }-\tau
_{1}^{\prime },\bar{q}_{1}+\bar{q}_{2}^{\prime }-\bar{\kappa}\right)  \notag\\
&&G_{ret}\left( \eta ^{\prime }-\tau _{2}^{\prime },-\bar{q}_{1}-\bar{q}_{2}^{\prime }\right) 
\left[ \bar{q}_{1}\cdot \left( \bar{\kappa}\times \bar{p}\right) \right] \left[ \bar{\kappa}\cdot 
\left( \bar{q}_{1}\times \bar{q}_{2}^{\prime }\right) \right]  \label{c-15} \\
&&\phi \left( \bar{q}_{1},\tau _{1}\right) \phi ^{\ast }\left( -\bar{q}_{1},\tau _{2}^{\prime }\right) 
\phi \left( \bar{\kappa}-\bar{q}_{1},\tau_{2}\right) \phi ^{\ast }\left( \bar{q}_{1}-\bar{\kappa},\tau _{1}^{\prime}\right)  \notag \\
&&\phi \left( \bar{p}-\bar{q}_{1},\tau _{1}\right) \phi ^{\ast }\left( \bar{q}_{1}-\bar{p},\tau _{2}\right) 
\phi \left( \bar{q}_{2}^{\prime },\tau_{1}^{\prime }\right) \phi ^{\ast }\left( -\bar{q}_{2},\tau _{2}^{\prime}\right) ,  \notag
\end{eqnarray}
for the square graph, which vanishes when the momenta integrals are peformed. 
\begin{eqnarray}
\xi_{c}\left( \kappa ,\eta ,\eta ^{\prime }\right)
&=&-e^{4}G_{ret}\left( \eta -\tau _{1},\bar{p}\right) G_{ret}\left( \eta -\tau _{2},\bar{\kappa}-\bar{p}\right) 
G_{ret}\left( \eta ^{\prime }-\tau_{1}^{\prime },\bar{q}_{1}-\bar{q}_{3}-\bar{p}\right)  \notag \\
&&G_{ret}\left( \eta ^{\prime }-\tau _{2}^{\prime },-\bar{\kappa}+\bar{p}-\bar{q}_{1}+\bar{q}_{3}\right)  \notag \\
&&\left[ \left( 2\bar{q}_{1}-\bar{p}\right) \cdot \left( \bar{\kappa}\times \bar{q}_{3}\right) -2\bar{q}_{1}\cdot 
\left( \bar{p}\times \bar{q}_{3}\right) \right] \left[ \left( \bar{p}-\bar{q}_{3}\right) \cdot 
\left( \bar{\kappa}\times \bar{q}_{1}\right) -2\bar{q}_{1}\cdot \left( \bar{p}\times \bar{q}_{3}\right) \right]  \nonumber \\
&&\phi \left( \bar{q}_{1},\tau _{1}\right) \phi ^{\ast }\left( -\bar{q}_{1},\tau _{2}^{\prime }\right) 
\phi \left( \bar{q}_{3},\tau _{2}\right)\phi ^{\ast }\left( -\bar{q}_{3},\tau _{1}^{\prime }\right)  \label{c-16} \\
&&\phi \left( \bar{p}-\bar{q}_{1},\tau _{1}\right) \phi ^{\ast }\left( \bar{q}_{1}-\bar{p},\tau _{1}^{\prime }\right) 
\phi \left( \bar{\kappa}-\bar{p}-\bar{q}_{3},\tau _{2}\right) \phi ^{\ast }\left( -\bar{\kappa}+\bar{p}+
\bar{q}_{3},\tau _{2}^{\prime }\right)  \notag
\end{eqnarray}
for the ``cross'' diagram and 
\begin{eqnarray}
\xi _{2b}\left( \kappa ,\eta ,\eta ^{\prime }\right)
&=&e^{4}G_{ret}\left( \eta -\tau _{1},p\right) G_{ret}\left( \eta -\tau
_{2},\kappa -p\right) G_{ret}\left( \eta ^{\prime }-\tau _{1}^{\prime
},-p\right) G_{ret}\left( \eta ^{\prime }-\tau _{2}^{\prime },p-\kappa
\right)  \notag \\
&&\left[ \left( 2\bar{q}_{1}-\bar{p}\right) \cdot \left( \kappa \times \bar{q
}_{3}\right) -2\bar{q}_{1}\cdot \left( \bar{p}\times \bar{q}_{3}\right) 
\right] ^{2}  \label{c-17} \\
&&\phi \left( \bar{q}_{1},\tau _{1}\right) \phi ^{\ast }\left( -\bar{q}
_{1},\tau _{1}^{\prime }\right) \phi \left( \bar{q}_{3},\tau _{2}\right)
\phi ^{\ast }\left( -\bar{q}_{3},\tau _{2}^{\prime }\right)  \notag \\
&&\phi \left( \bar{p}-\bar{q}_{1},\tau _{1}\right) \phi ^{\ast }\left( \bar{q
}_{1}-\bar{p},\tau _{1}^{\prime }\right) \phi \left( \bar{\kappa}-\bar{p}-
\bar{q}_{3},\tau _{2}\right) \phi ^{\ast }\left( p+\bar{q}_{3}-\bar{\kappa}
,\tau _{2}^{\prime }\right)  \notag
\end{eqnarray}
for the ``two bubbles'' figure.
It is clear that for a given volume, the flux of magnetic field across its boundary is not zero; in other words, the boundary of
$\mathcal{V}$ is not a magnetic surface. In principle, this fact renders $\mathcal{H}_{M}$ as defined in (\ref{b-1}), 
a gauge dependent quantity. As stated above, Berger and Field \cite{berger} gave a gauge-independent
measure of magnetic helicity suitable for this situation, based on the difference between the magnetic helicities of 
two field configurations that have a common extension outside the considered volume. The only constraint that 
definition must satisfy is that the sources of the fields must be bounded in order to guarantee that the surface at infinity 
is a magnetic one, which is equivalent to say that the magnetic field at infinity must vanish. The fact that in our model 
$B\rightarrow 0$ at infinity \cite{ckm-98} allows us to consider the boundary at infinity as a magnetic surface, in spite of the
fact that the stochastic currents exist in the whole space (cosmological particle creation is not restricted to the particle horizon 
\cite{birrel-94}). Below, in section \ref{gd} we discuss how to apply Berger's criterion to our fields.

\subsection{\label{dim}Some dimensions}

In this subsection we provide some (dimensional) quantities that will be used later to obtain concrete values of magnetic intensities. 
We are interested in two scales: one is large, e.g., of the order of the galactic commoving scale (or larger) for which, due to the fact that 
it remains outside the particle horizon during most of Radiation dominance, we can consider diffusive evolution of the magnetic field. 
The other is of the order of the horizon at reheating, where the magnetic field may be subjected to an inverse cascade of 
magnetic helicity if the medium is turbulent. In view of the fact that Reheating happens in a very short period of time, we can consider 
that during it the particle horizon remains practically constant and equal to the one in Inflation, i.e., $h_{rh} \sim 1$. 
When the universe enters into the radiation dominated regime, the horizon grows as $h \sim \eta^2$, a fact that in terms of the 
temperatures can be expressed as $h\left( T\right) \sim \left( T_{rh}/T\right)^2 h_{rh}$. In the presence of turbulence, the viscous 
dissipation scale is usually estimated as $\lambda_{diss} \simeq h/R_e$ with $R_e$ the Reynolds number. According to
Ref. \cite{cal-kan-10}, during Reheating can be taken as $R_e\sim 100$ and smaller for later epochs \cite{giov-11}.
Concerning the galactic size we have that today a galactic commoving scale (i.e., a not-collapsed  scale that contains all the matter 
of a Milky Way-like galaxy) is $\lambda_G \simeq 1$ Mpc. For the Hubble constant during Inflation we consider $H=10^{11}-10^{13}~GeV$; 
the mass of the scalar field can be taken as $m\simeq 100~GeV$; we also have $1GeV = 1.9733 \times 10^{-14}$ cm; the electric 
conductivity takes the usual form considered in the literature, $\sigma_{0}\simeq \left( T_{rh }/e^{2}\right) 
\left[Hm_{pl}/T_{rh }^{2}\right] ^{1/4}\simeq 1.37~\times 10^{19/2}GeV$, $e^{2}=1/137$, where we considered 
$T_{rh }\simeq 10^{2}~GeV$ and $m_{pl}=10^{19}~GeV$.  Finally, the temperature at matter-radiation equilibrium can be 
taken as $T_{eq}\sim 1$ eV. Thus the different quantities that appear in the formulae lay in te intervals
\begin{eqnarray}
\frac{m}{H} &\simeq &10^{-11}-10^{-9}  \label{c-151} \\
\kappa _{gal} &=&\frac{k_{gal}^{phys}}{H}\simeq 10^{-51}-10^{-49}
\label{c-161} \\
\frac{\sigma _{0}}{H} &\simeq &10^{-7/2}-10^{-3/2}  \label{c-171}
\end{eqnarray}

\subsection{\label{hm-tir}Magnetic Helicity due to the Transition Inflation-Reheating}

When the transition from Inflation to Reheating occurs, the vacuum state of
the fields during Inflation turns into a particle state \cite{birrel-94}.
Mathematically, this means that the positive frequency modes of the field in
the inflationary epoch can be expressed as a linear combination of positive
and negative frequency modes in Reheating.
As we are dealing with a charged scalar field, electric charges will appear
and, as there are equal numbers of positive and negative carriers the mean
value of the induced electric current is zero, but not its rms deviation,
which will source the stochastic magnetic field. Therefore in order to
evaluate $\xi \left( \kappa ,\eta ,\eta ^{\prime }\right) $ we
express the $\phi \left( p,\tau \right) $ fields of Inflation in terms of
the ones in reheating by the usual Bogoliubov transformation \cite{birrel-94}:
\begin{equation}
\phi \left( p,\tau \right) =\alpha _{p}\phi _{R}\left( p,\tau \right) +\beta
_{p}\phi _{R}^{\ast }\left( p,\tau \right)  \label{d-1}
\end{equation}%
$\alpha _{p}$ and $\beta _{p}$ being the Bogoliubov coefficients satisfying
the normalization condition $\left\vert \alpha _{p}\right\vert^{2}-\left\vert \beta _{p}\right\vert ^{2}=1$. 
The different field products then read
\begin{eqnarray}
\phi \left( p,\tau \right) \phi ^{\ast }\left( -p,\tau ^{\prime }\right)
&=&\phi _{R}\left( p,\tau \right) \phi _{R}^{\ast }\left( -p,\tau ^{\prime
}\right) +\beta _{-p}^{\ast }\alpha _{p}\phi _{R}\left( p,\tau \right) \phi
_{R}\left( -p,\tau ^{\prime }\right)  \notag \\
&&+\beta _{p}\alpha _{-p}^{\ast }\phi _{R}^{\ast }\left( p,\tau \right) \phi
_{R}^{\ast }\left( -p,\tau ^{\prime }\right)  \label{d-2} \\
&&+\left\vert \beta _{p}\right\vert ^{2}\left[ \phi _{R}\left( p,\tau
\right) \phi _{R}^{\ast }\left( -p,\tau ^{\prime }\right) +\phi _{R}^{\ast
}\left( p,\tau \right) \phi _{R}\left( -p,\tau ^{\prime }\right) \right] 
\notag
\end{eqnarray}
The first term represents the vacuum-to-vacuum transition, the second and
third account for a mixing of positive and negative frequency modes, while
the fourth, proportional to $\left\vert\beta_{p}\right\vert ^{2}$, is due 
to solely the negative frequency modes, i.e. to the transition vacuum to 
particle state. As the main contribution comes from this term from now on we consider only it.

In a non-instantaneous transition the creation of  small scale modes depends 
on the details of the transition, while for superhorizon modes details of the 
transition do not matter. For subhorizon modes we have \cite{cal-kan-10}
\begin{equation}
\beta_p \sim\frac{i}{16\tau_0 p^5}, \qquad p \geq 1 \label{d-3a}
\end{equation}
while for $p<1$ this coefficient reads (see Appendix \ref{apa})
\begin{equation}
\beta_p \simeq -i\sqrt{\frac{H}{2m}}\sqrt{\frac{9}{8}}p^{-3/2}, \qquad p < m/H \label{d-4a}
\end{equation}
\begin{equation}
\beta_p \simeq -i\frac{3}{8}e^{-ip}p^{-3}, \qquad m/H < p < 1 \label{d-5a}
\end{equation}
We see that the contribution of subhorizon modes to the magnetic helicity correlation
function will be suppressed relative to the one of superhorizon ones. However, according 
to  the results of Ref. \cite{cal-kan-10} subhorizon fluctuations are responsible for 
a mildly turbulent flow on scales of the order of the horizon size, with Reynolds numbers 
$R_e \simeq 100$. 

An important comment about the infrared limit in expr. (\ref{d-4a}) is in order. It is clear that
that expression blows out for $p\rightarrow 0$. This is due to the approximations made to solve 
Klein-Gordon equation during Reheating. To give a physical lower limit we must consider the largest
homogenous patch created during Inflation as the largest possible scale, which according to Refs. 
\cite{cal-sake-92,cal-sake-93} can be considered as about 10 times the horizon during Inflation.

\subsection{\label{gd}Gauge Dependence}

In principle, the dependence of the rms value of $\mathcal{H}_M$ on the gauge could be
analised by adding to $\mathcal{A}\left(\bar p\right)$ a term of the form $ip_j\psi
\left(p\right)$ to the r.h.s. of expression \ref{c-5}, $\psi\left(p\right)$ being 
an arbitrary scalar function. Then in $\Xi\left(\bar\kappa , \bar\kappa^{\prime}, 
\eta ,\eta^{\prime}\right)$ there will appear a term proportional to $\left[\mathcal{B}_i
\left(\bar\kappa - \bar p,\eta\right) p^i\right] \left[\mathcal{B}_j\left(\bar\kappa^{\prime} - 
\bar p^{\prime},\eta^{\prime}\right) p^{\prime j}\right]$ that cancels indentically 
when the angular integrals are performed. This means that only ``on average'' our result 
is gauge invariant. If we consider a gauge term like $\mathcal{G}=\mathcal{B}_{ij}p^i p^j$,
with $\mathcal{B}_{ij}$ an arbitrary magnetic correlation function, generally it will satisfy that
$\langle\mathcal{G}^2\rangle \not=0$, and consequently the associated magnetic helicity
will not be statistically gauge independent.

We then reason according to Berger and Field as follows. For any volume 
$\mathcal{V}$  we were interested in (a galaxy, a galaxy cluster, horizon at a certain 
epoch, etc),  its surrounding region corresponds to the rest of the embedding space whose 
characteristic scale  is very much larger than  $\mathcal{V}^{1/3}$. If the magnetic 
correlation tends to zero from a certain scale on, i.e., if  for $\kappa \rightarrow 0$
it is proportional to a power of $\kappa$, then for an observer inside $\mathcal{V}$ the
field outside is statistically equivalent to a vanishing field. Consequently we could interpret $\Xi\left(\mathcal{V},
\eta ,\eta^{\prime}\right)$  as the difference between two magnetic field configurations: 
one being the calculated through the diagrams above, and the other a null configuration.

To quantify this assertion, we begin by noting that the trace of the magnetic 
field correlation function is given by 
\begin{eqnarray}
\langle \mathcal{B}_i\left(\bar \kappa,\eta\right) \mathcal{B}_i\left(\bar\kappa^{\prime},
\eta^{\prime}\right)
&=& e^2 G_{ret}\left(\bar\kappa,\eta - \tau_1\right) G_{ret}\left(\bar\kappa^{\prime},
\eta^{\prime} -\tau_2\right)\nonumber\\
&&\delta\left(\bar\kappa - \bar p_1\right)\delta\left(\bar q_1 + \bar q_2 - \bar p_1\right)
\delta\left(\bar\kappa^{\prime} - \bar p_2\right)\delta\left(\bar q_1^{\prime} +\bar q_2^{\prime}
-\bar p_2\right)\label{g-1}\\
&&\epsilon_{ils}\epsilon_{ijm}p_l\left(q_{1s} - q_{2s}\right)\left(\kappa^{\prime}_j\right)
\left(q_{1m}^{\prime} - q_{2m}^{\prime}\right)\nonumber\\
&&\langle \phi^1\left(\bar q_1,\tau_1\right)\phi^1\left(\bar q_1^{\prime},\tau_2\right)\rangle
\langle\phi^2\left(\bar q_2,\tau_1\right)\phi^2\left(\bar q_2^{\prime},\tau_2\right)\rangle
\nonumber
\end{eqnarray}
where we again assume integration over $\tau_i$ and over repeated momenta other than $\kappa$. 
Using again decomposition (\ref{c-10}) for the scalar fields we have that from each mean value the
only non-null contribution is
\begin{equation}
\langle \phi^i\left(\bar q,\tau\right)\phi^i\left(\bar q^{\prime},\tau^{\prime}\right)\rangle
\rightarrow \phi\left(\bar q,\tau\right)\phi^{\ast}\left(\bar q^{\prime},\tau^{\prime}\right)
\delta\left(\bar q + \bar q^{\prime}\right) \label{g-2}
\end{equation}
The integration of the delta functions in eq. (\ref{g-1}) produce again a $\delta\left(\bar\kappa + \bar\kappa^{\prime}\right)$,  
i.e. momentum conservation. Writing the trace of the magnetic field correlation as
$\langle \mathcal{B}_i\left(\bar \kappa,\eta\right) \mathcal{B}_i\left(\bar\kappa^{\prime},\eta^{\prime}\right)\rangle = 
\cal{M}\left(\bar\kappa ,\eta ,\eta^{\prime}\right) \delta\left(\bar\kappa + \bar\kappa^{\prime}\right)$
we obtain the following expression for the prefactor for trace of the magnetic field correlation function
\begin{eqnarray}
\mathcal{M}\left(\bar \kappa,\eta ,\eta^{\prime}\right) &=& 4e^2 G_{ret}\left(\bar\kappa,\eta - \tau_1\right) G_{ret}\left(-\bar\kappa,
\eta^{\prime} -\tau_2\right)\label{g-3}\\
&&\left[\bar\kappa\times\bar q_1\right]^2
\phi\left(\bar q_1,\tau_1\right)\phi^{\ast}\left(-\bar q_1,\tau_2\right)
\phi\left(\bar\kappa - \bar q_1,\tau_1\right)\phi^{\ast}\left(\bar q_1-\bar\kappa ,\tau_2\right)
\nonumber
\end{eqnarray}

As above, for any considered volume $\mathcal{V}$ we are interested in, its surrounding region, $\mathcal{V}_s$ corresponds to the 
rest of the  embedding space and satisfies $\mathcal{V} \ll \mathcal{V}_s$. The magnetic field configuration in those  regions corresponds to 
superhorizon  modes whose evolution is in general diffusive. 
Therefore to estimate the rms magnetic  intensity at a certain time on a comoving scale 
$\kappa \ll 1$ we use expressions (\ref{apb-4}) for the  retarded propagators, (\ref{apa-7b}) 
for the modes and (\ref{d-4a}) for the Bogoliubov coefficients. We then write
\begin{eqnarray}
\mathcal{M}\left(\bar \kappa,\eta ,\eta^{\prime}\right) &\sim&  4e^2\left(\frac{H}{\sigma_0}\right)^2\left(\frac{H}{m}\right)^2
\frac{\left\vert\bar\kappa\times\bar q_1\right\vert^2}{q_1^3\left\vert\bar\kappa - \bar q_1\right\vert^3}
\nonumber\\
&\times & \int_0^{\tau_f} d\tau \int_0^{\tau_f} d\tau^{\prime}
\frac{\left[1-e^{-\sigma_0\left(\eta - \tau\right)/H}\right]}{\left(1+\tau/2\right)^2}
\frac{\left[1-e^{-\sigma_0\left(\eta^{\prime} - \tau^{\prime}\right)/H}\right]}{\left(1+
\tau^{\prime}/2\right)^2}\label{g-4}
\end{eqnarray}
where $\tau_f$ is the lifetime of the stochastic electric current. As $\tau_f \ll 1$, it can be
neglected in the denominator of each of the time integrals, which can then be estimated as
\begin{equation}
\int_0^{\tau_f} d\tau\left[1-e^{-\sigma_0\left(\eta - \tau\right)/H}\right]
\simeq \tau_f \left[ 1 + e^{-\sigma_0\eta / H}\right]
\end{equation}
We can estimate the order of magnitude of the integral in the momenta by counting powers. 
In the numerator we have 5 powers of $\kappa$ and 5 powers of $q_1$ (three from the integration
meassure and two from the square of the cross product). In the denominator there are 3 powers of
$q_1$ and 3 powers of $\kappa$. Consequently the overall contribution of the integrals in the momenta
is a factor $\kappa^2 q_1^2$. To estimate their numerical value we must filter out the length scales smaller
than those corresponding to $\kappa_s = 2\pi \mathcal{V}_s^{-1/3}$ because they oscillate inside $\mathcal{V}_s$,
and so the main contribution from the momentum integrals is 
$\sim \kappa_s^4$. Finally, taking the coincidence limit $\eta = \eta^{\prime}$ we have
\begin{equation}
\langle \mathcal{B}_{\mathcal{V}_s}^2\rangle \sim  e^2\left(\frac{H}{\sigma_0}\right)^2
\left(\frac{H}{m}\right)^2
\tau_f^2 \left[ 1 + e^{-\sigma_0\eta / H}\right]^2\kappa_s^4\label{g-5}
\end{equation}
Thus we see that for $\kappa_s\rightarrow 0$, i.e., for large volumes as those outside the considered structure, 
the magnetic field vanishes and so does their associated magnetic helicity. Therefore we can interpret our expression for the 
rms value of the magnetic helicity in a given volume $\mathcal{V}$ in the spirit of the work by Berger and Field, as the substraction 
of two magnetic configurations, one of them being effectively zero.

\subsection{\label{mhls} Magnetic Helicity on Large Scales}

In this section we evaluate the magnetic helicity on large scales due to both smooth and fluctuating fields on
that scale. From the form of the Bogoliubov coefficients, eqs. (\ref{d-3a})-(\ref{d-5a}) we see
that the main contribution is due to the modes with $p < m/H$, i.e., eq. (\ref{d-4a}). In this case the modes
are given by eq. (\ref{apa-7b}), whereby
\begin{equation}
\phi_R\left(p,\tau\right)\phi_R^{\ast}\left( -p,\tau^{\prime}\right) + c.c \sim \frac{6}{\pi}
\frac{\cos\left\{\left(2m/3H\right)\left[\left( 1+\tau /2\right)^{3}-
\left( 1+\tau^{\prime }/2\right)^{3}\right]\right\}}{
\left( 1+\tau /2\right)\left( 1+\tau^{\prime }/2\right)}\label{h-1}
\end{equation}
We are interested in the value of the magnetic helicity at present time. In this limit, and for small momenta, the retarded 
propagators are given by eq. (\ref{apb-5}), namely $G_{ret}\left(p,\eta ,\tau\right)
\sim H/\sigma_0$. The lifetime of the electric currents
was calculated in Ref. \cite{cal-kan-10} in physical time (see eq. (4.9) of that reference). In conformal (dimensionless) time
it reads $\tau_f \simeq \left( Ht_f\right)^{1/3}$ and from the reference mentioned just above we have
\begin{equation}
\tau_f\left( p\right) \sim \left(\frac{m}{H}\right)^{1/3} \frac{\tau_0^{2/3}}{\pi^{1/3}e^{4/3}}
\frac{\left[p^2 + \left(m/H\right)^2 \right]^{1/6}}{\left[\left(3/2\right)\left(H/m\right)^3\tau_0^2
+ \left( 9/16\right)^4 \right]^{1/3}}\label{h-2}
\end{equation}
with $\tau_0$ the duration of the transition Inflation-Reheating and where we considered that $a\left(\tau\right) \sim 1$
around that transition. Considering the values for $m/H$ quoted in Section \ref{dim} and the smallest possible value for
$\tau_0$, namely the Planck time, it is seen that $\left(H/m\right)^3\tau_0^2\gg 1$, and for $p < m/H$
expr. (\ref{h-2}) gives
\begin{equation}
\tau_f\left( p\right) \sim \left(\frac{m}{H}\right)^{5/3}\frac{1}{\pi^{1/3}e^{4/3}}\label{h-3}
\end{equation}
Observe that $\tau_f\ll 1$. This fact allows us to neglect $\tau ,~\tau^{\prime}$ in front of 1 in expr. (\ref{h-1}) 
reducing the time integrals in each of the non-trivial graph to simply $\int_0^{\tau_f}d\tau = \tau_f$. This simplification
permits to write the sum $\Xi = \Xi_c + \Xi_{2b}$ as
\begin{equation}
\xi\left(\kappa ,\eta , \eta^{\prime}\right)\sim
8e^4 \left(\frac{H}{\sigma_0}\right)^4\left(\frac{6}{\pi}\right)^4 \tau_f^4 \left\vert\beta_{q_1}\right\vert^2
\left\vert\beta_{q_3}\right\vert^2\left\vert\beta_{q_1-p}\right\vert^2\left\vert\beta_{\kappa-p-q_3}\right\vert^2 
B\left[C - B\right]\label{h-4}
\end{equation}
with
\begin{eqnarray}
B &=& \left( 2\bar q_1 - \bar p\right)\cdot\left(\bar\kappa\times\bar q_3\right) - 2\bar q_1\cdot\left(\bar p\times\bar q_3\right)
\label{h-5}\\
c &=& \left( \bar p - \bar q_3\right)\cdot\left(\bar\kappa\times\bar q_1\right) - 2\bar q_1\cdot\left(\bar p\times\bar q_3\right)
\label{h-6}
\end{eqnarray}
The factor $B\left[C - B\right]$ was solved in Appendix \ref{apc} giving the non-null result $B\left[C - B\right]= \left\{\bar\kappa\cdot
\left[\bar q\times\left(\bar q_1 + \bar q_3\right)\right]\right\}^2$, with $\bar q = \bar p - \bar q_1$. Replacing expression (\ref{d-4a}) 
for the Bogoliubov coefficients and (\ref{h-3}) for $\tau_f$, the magnetic helicity correlation function (\ref{h-4}) for large scales and 
long times becomes
\begin{equation}
\xi\left(\kappa\right)\sim 8 \left(\frac{H}{\sigma_0}\right)^4\left(\frac{m}{H}\right)^{8/3}
\frac{6^4}{\left(\pi^4 e\right)^{4/3}} \frac{\left\{\bar\kappa\cdot\left[\bar q\times\left(\bar q_1 + \bar q_3\right)\right]
\right\}^2}{q_1^3q_3^3q^3\vert \kappa - \bar q - \bar q_1 - \bar q_3\vert^3}\label{h-7}
\end{equation}
There remain the integrations over the momenta other than $\kappa$. As before, their contribution can be roughly estimated by counting
powers as before. Since we are considering scales such that $\kappa \ll p < m/H$, we can disregard $\kappa$ in the denominator 
of expr. (\ref{h-7}). 
Then we have that the power of the scalar field momenta in the numerator is 13 (nine from the integration
meassures plus 4 from the square of the cross product) while in the denominator their power is 12 (three from each of the four
Bogoliubov factor), giving as result a factor of the form $p_{max}$. As the maximum value allowed by the approximations is 
$p_{max}\sim m/H$ the result is
\begin{equation}
\xi\left(\kappa\right)\sim  8\left(\frac{H}{\sigma_0}\right)^4\left(\frac{m}{H}\right)^{11/3}
\frac{6^4}{\left(\pi^4 e\right)^{4/3}} \kappa^2\label{h-8}
\end{equation}
According to definition (\ref{c-1}) there remain the integration over $\kappa$'s, wich gives an extra factor $\kappa^3$, and the two integrations over the volume, which according to expr. (\ref{c-4}) each one gives a factor of $\kappa^{-3}$. Then the contribution
of all non-null graphs to the magnetic helicity is
\begin{equation}
\Xi\left(\kappa\right)\sim  8\left(\frac{H}{\sigma_0}\right)^4\left(\frac{m}{H}\right)^{11/3}
\frac{6^4}{\left(\pi^4 e\right)^{4/3}} \kappa^{-1}\label{h-9}
\end{equation}
Finally, the rms value of the magnetic helicity in a comoving volume $\kappa^{-3}$ can be considered as simply 
the squareroot of expr. (\ref{h-9}) giving
\begin{equation}
\mathcal{H}_M\left(\kappa\right)\sim  \left(\frac{H}{\sigma_0}\right)^2\left(\frac{m}{H}\right)^{11/6}
\frac{102}{\left(\pi^4 e\right)^{2/3}} \kappa^{-1/2}\label{h-10}
\end{equation}
Note that this helicity corresponds to regular as well as irregular fields on the considered volume, i.e. to the total magnetic
field due to fluctuations up to the scale $m/H$. The density of defects on a scale $\kappa$ is found by multiplying 
expr. (\ref{h-10}) by $\kappa^3$, finding that it varies as $\kappa^{5/2}$. Using the fact that a fractal of dimension $D$
embedded in a spherical volume has a number density that scales with the radius $\kappa^{-1}$ from an occupied point as
$\Gamma\left(\kappa\right)\propto \kappa^{3-D}$
we estimate the fractal dimension of the distribution of topological defects as being $D=3-5/2 = 1/2$.

\subsection{\label{hmgal}Magnetic Helicity due to smooth fields}

In this section we evaluate the magnetic helicity and its corresponding magnetic field on large scales due to smooth fields only.
The estimates will be the values that they would have today if they were not affected by the galaxy formation process. 
The expression we are seeking for is obtained directly from eq. (\ref{h-7}), estimating the integrals in the momenta other than 
$\kappa$ by filtering the frequencies higher than the one associated to the considered scale, say $\kappa_{\lambda}$. We obtain 
\begin{equation}
\Xi \left( \kappa_{\lambda}\right) \sim e^4\left(\frac{H}{m}\right)^4\left(\frac{H}{\sigma_0}\right)^4
\tau_{f}^{4}  \label{d-1b}
\end{equation}
i.e., a scale independent number. As before, we estimate the magnetic helicity on the considered scale by simply taking the
squareroot of expression (\ref{d-1b}) and, after replacing $\tau _{f}$ from (\ref{h-3}) we obtain
\begin{equation}
\mathcal{H}_{M}\left( \kappa_{\lambda}\right) \sim  e^{-2/3}\left(\frac{m}{H}\right)^{4/3}
\left(\frac{H}{\sigma_0}\right)^2\label{d-2b}
\end{equation}
Using the numbers given by (\ref{c-151})-(\ref{c-171}) we have on a
galactic scale  
\begin{equation}
H_{M}\left( \kappa _{G}\right) \sim 10^{-11}-10^{-9}  \label{d-3}
\end{equation}
which is a very small number. From eq. (\ref{d-3}) we obtain the number density of links of the smooth field by multiplying
by $\kappa_s^3$, obtaining $e^{-2/3}\left(\frac{m}{H}\right)^{4/3}\left(\frac{H}{\sigma_0}\right)^2\kappa_s^3$, which means that
the fractal dimension of the distribution is $D=0$. This would mean that for large scales the number of defects becomes independent of
the scale.
It is interesting to estimate the order of magnitude of the coherent magnetic field associated to this helicity on a galactic scale. 
We can crudely do that by taking $\mathcal{A}.\mathcal{B}\sim \mathcal{H}_{M}\kappa _{G}^{3}$ with ${\cal A}\sim {\cal B}/\kappa _{G}$, 
whereby $\mathcal{A}.\mathcal{B}\sim \mathcal{B}^{2}/\kappa _{G}\sim \mathcal{H}_{M}\kappa _{G}^{3}\rightarrow \mathcal{B}\sim 
\mathcal{H}_{M}^{1/2}\kappa _{G}^{2}$, i.e.
\begin{equation}
\mathcal{B}\sim e^{-1/3}\left(\frac{m}{H}\right)^{2/3}\left(\frac{H}{\sigma_0}\right)\kappa _{G}^{2} \label{d-4}
\end{equation}
Recalling that the physical field is $B = H^2\mathcal{B}$ and that 1 GeV$^2 \simeq 10^{20}$ Gauss we obtain that on a galactic scale 
the helical fields have an intensity of the order
\begin{equation}
B_G^{hel} \sim 10^{-61} ~\mathrm{Gauss} \label{d-5}
\end{equation}
which is a very small value, that however agrees with previous estimates in the literature \cite{ckm-98,giov-shap-00}. Observe however 
that expr. (\ref{d-4}) coincides with the squareroot of expr. (\ref{g-5}) when $\tau_f$ is replaced by expr. (\ref{h-3}) and the limit 
$\eta\rightarrow\infty$ is taken. This means that the magnetic fields smooth on large scales induced by the mechanism considered in this 
paper are indeed helical.

\subsection{\label{hm-ss}Magnetic Helicity at Small Scales}

The important feature of magnetic helicity for large scale magnetogenesis is the fact that it performs an inverse cascade 
if the medium where it evolves is turbulent. We shall then make the hypothesis that after reheating the plasma is in a state of 
decaying turbulence, where self-organization of magnetic structures can happen. The intensity of the turbulence is determined by the 
Reynolds number, which during reheating was estimated in Ref. \cite{cal-kan-10} to be $R_e \sim 100$, while for times around 
electron-positron annihilation was calculated to be $R_e\sim 0.03$ \cite{giov-11}. Therefore a decaying turbulence scenario in the early
universe is possible, the turbulence being mild.

Turbulence is a causal phenomenon, i.e., it happens on scales equal or smaller than the particle horizon. 
We must evaluate the intensity of magnetic helicity in regions of size of at most the 
particle horizon at the time $\tau_f$, when the sources of magnetic field vanish. After that moment the evolution of the field intensity and coherence scale will be considered to be due to the inverse cascade of magnetic helicity. We shall analize that evolution in the light
of the simple model discussed in Ref. \cite{biskamp-03} (see also Ref. \cite{kahniashvili-13}), whereby 
the comoving coherence length of the magnetic field grows as $\ell \propto \ell_0\eta^{2/3}$ and the total comoving energy 
$\mathcal{E}_M$ contained in a given volume decays as $\mathcal{E}_M \propto \mathcal{E}_M(0)\eta^{-2/3}$. Moreover, when the 
volume is fixed that law can be applied to the comoving magnetic energy density, i.e., we can consider 
$\mathcal{B}^2 \propto\mathcal{B}(0)\eta^{-2/3}$. 
Those laws arise from the conservation of magnetic helicity during the process of inverse cascade. Observe that the comoving coherence 
length remains always smaller than the comoving horizon, which after Inflation grows as $\propto\eta$ (the physical horizon growing as 
$\propto\eta^2$).

We begin by estimating the dimensionless time interval elapsed between the end of reheating and matter-radiation equilibrium. 
Knowing that the horizon grows as $\eta^2$, that interval can be estimated as $\Delta_{r-e} = \sqrt{h_{eq}/h_{rh}}$. For the horizon at 
matter-radiation equilibrium 
we take it as  \cite{kolb-tur} $\simeq 11~ \mathrm{Mpc}\simeq 16\times 10^{38}~\mathrm{GeV}^{-1}$. For reheating, we assume
that the transition Inflation-Reheating was fast enough, and that the electric currents vanished in a very short time, $\tau_f$,
to assume that the horizon at the end of reheating is not very different from $H^{-1}$, with $H$ the Hubble constant during Inflation.
Therefore $\Delta_{r-e} = \sqrt{16\times 10^{38}~\mathrm{GeV}^{-1}H}$. Using the values for $H$ quoted in Section \ref{dim} that 
period of time would be in the interval $4\times 10^{25} \lesssim \Delta_{r-e}\lesssim 4\times 10^{26}$.

To evaluate the magnetic helicity correlation we firstly note that the modes that will contribute the most 
are those such that $q_i,~q < 1$. This is due to the form of the Bogoliubov coefficients (\ref{d-3})-(\ref{d-5}).
Because of the different intervals used to find the scalar field modes and the corresponding Bogoliubov
coefficients, we must consider two intervals: $q,~q_i \to \left( \Lambda_{ir}, m/H\right)$, with Bogoliubov
coefficient given by expr. (\ref{d-4}) and $q,~q_i \to \left( \sim m/H, \sigma_0/H\right)$ with Bogoliubov coefficient 
given by expr. (\ref{d-3}). In the first interval $\Lambda_{ir}$ is an infrared cut-off that corresponds to the scal
of the largest inflationary patch discussed above. In the second interval, the upper limit $\sigma_0/H$ is considered for
consistency with the approximations made in Appendix \ref{apb} to find the retarded propagator for large scales, eq. (\ref{apb-4}).
With these considerations in the momenta and for the short time periods we consider, we can disregard the time
dependence in the corresponding mode functions, eqs. (\ref{apa-7b}) and (\ref{apa-10}) because
$1\gg \tau_f$. Therefore the mode functions during reheating to be used to evaluate the integrals in the
momenta read $\phi_p \sim \sqrt{6/\pi}$ for $p < m/H$ and $\phi_p \sim 1/2^{1/2} p^{-3/2}$ for $m/H \ll p < \sigma_0/H$.
The contribution of the ``cross'' plus the ``two bubbles'' diagrams is then similar to the corresponding to
large scales, namely
\begin{eqnarray}
\xi_{2b}+\xi_c &\sim& -e^4 G_{ret}\left(\eta ,\tau_1\right)G_{ret}\left(\eta ,\tau_2\right)
G_{ret}\left(\eta^{\prime} ,\tau_1^{\prime}\right)G_{ret}\left(\eta^{\prime} ,\tau_2^{\prime}\right)\nonumber\\
&&\left\{\bar\kappa\cdot\left[\bar q\times \left(\bar q_1 + \bar q_3\right)\right]\right\}^2 
\vert\beta_{q_1}\vert^2 \vert\beta_{q_3}\vert^2\vert\beta_{q-q_1}\vert^2\vert\beta_{\kappa - q -q_3}\vert^2\label{e-1}\\
&&G^{+}\left(q_1,\tau_1,\tau_1^{\prime}\right)G^{+}\left(q_3,\tau_2,\tau_2^{\prime}\right)
G^{+}\left(q-q_1,\tau_1,\tau_1^{\prime}\right)G^{+}\left(\kappa - q - q_3,\tau_2,\tau_2^{\prime}\right)\nonumber
\end{eqnarray}
with $G^{+}\left(q,\tau,\tau^{\prime}\right) = \phi\left(q,\tau\right)\phi^{\ast}\left(-q,\tau^{\prime}\right)
-\phi^{\ast}\left(q,\tau\right)\phi\left(-q,\tau^{\prime}\right)$.
As the time dependence of the momenta has dissappeared, the time integrals involve only the expressions for the
retarded propagators. For $p < \sigma_0/H$ and time intervals $\Delta\tau \ll 1$ they
are given by expr. (\ref{apb-6}) below, i.e., $G_{ret}\left(p ,\eta ,\tau \right)\simeq \left(\eta - \tau\right)
\Theta\left(\eta - \tau\right)$. The integration between $\tau = 0$ and $\tau = \tau_f$ is then straightforward,
giving each propagator a contribution of $\tau_f^2$, the overall time contribution thus being $\tau_f^8$. Replacing
the corresponding expressions for the Bogoliubov coefficients and the modes, eq. (\ref{e-1}) can be written as
$\xi = \xi_1 + \xi_2$ with $\xi_1$ corresponding to the momentum interval $p < m/H$ and $\xi_2$ to $m/H < p < \sigma_0/H$.
Explicitly we have
\begin{equation}
\xi_1 \sim e^4\tau_f^8 \left(\frac{6}{\pi}\right)^4\left(\frac{H}{m}\right)^4
\frac{\left\{\bar\kappa\cdot\left[\bar q\times \left(\bar q_1 + \bar q_3\right)\right]\right\}
^2}{q_1^3q_3^3q^3\vert \bar\kappa - \bar q - \bar q_1 - \bar q_3\vert^3}\label{e-2}
\end{equation}
and
\begin{equation}
\xi_2 \sim e^4\tau_f^8 
\frac{\left\{\bar\kappa\cdot\left[\bar q\times \left(\bar q_1 + \bar q_3\right)\right]\right\}^2}{q_1^9q_3^9q^9
\vert \bar\kappa - \bar q - \bar q_1 - \bar q_3\vert^9}\label{e-3}
\end{equation}
To roughly estimate all momentum integrals we again count powers. As $\kappa \sim 1$ we begin 
by neglecting the $q$'s in front of $\kappa$ in both expressions. For $\Xi_1$ we obtain a contribution of the form
$\sim \kappa^2 q^2 /q_1^2$ and considering the contribution that gives the largest value, namely $q,~q_1\sim m/H$ we obtain
\begin{equation}
\Xi_1 \sim e^4\tau_f^8 \left(\frac{6}{\pi}\right)^4 \kappa^2\label{e-4}
\end{equation}
For $\Xi_2$ the counting of powers give a factor of the form $\kappa^{-4} q_1^{-4}q^{-4}q_3^{-6}$. 
and in the corresponding interval $(m/H,\sigma_0/H)$ we consider the contribution of the lower limit, i.e. again
$m/H$, as it gives the largest value. Te result is
\begin{equation}
\Xi_2 \sim e^4\tau_f^8 \left(\frac{H}{m}\right)^{14}\frac{1}{\kappa^4}\label{e-5}
\end{equation}
An important comment is in order: according to the approximations made to solve the Klein-Gordon equation during Reheating 
in Appendix \ref{apa}, the intervals $p< m/H$ and $p> m/H$ determined the two different 
sets of solutions that were used
to calculate the Bogoliubov coefficients (\ref{d-4}) and (\ref{d-5}). Taking the limit $p\rightarrow m/H$ in both of those
expressions, we see that they are not continuous in that limit, which means that the mentioned approximations break down. Consequently both 
$\Xi_1$ and $\Xi_2$ found using that value of momentum must be considered as upper bounds to the possible realistic values.
We see that the main contribution to the rms value of the magnetic helicity is due to $\Xi_2$, which after replacing the
expression (\ref{h-3}) reads
\begin{equation}
\Xi_2 \sim \frac{1}{\pi^{8/3}e^{20/3}} \left(\frac{H}{m}\right)^{2}\frac{1}{\kappa^4}\label{e-6}
\end{equation}
We estimate the magnetic helicity on a volume $\kappa^{-3}$ again by taking the squareroot of (\ref{e-6}), and thus see that the
number of links scales as
\begin{equation}
\mathcal{H}_M \sim \frac{1}{\pi^{4/3}e^{10/3}} \frac{H}{m}\kappa^{-2}\label{e-7}
\end{equation}
and their density as $\propto\kappa$. The associated fractal dimension is $D=2$.

We now turn our attention to the evolution of the magnetic field from its value induced at the end of reheating in a scale of 
the order of the horizon up to matter-radiation equilibrium. That evolution will be due to a possible turbulent state during radiation 
dominance, which would allow an inverse cascade of magnetic helicity. We thus estimate again the comoving magnetic energy density as
$\mathcal{B}^2_0\sim \mathcal{H}_{M,0}\kappa^{4}$, and for $\kappa \simeq 1$ it is simply $\mathcal{B}_0^2\sim \mathcal{H}_{M,0}$.
At matter-radiation equilibrium we would then have a comoving magnetic field intensity of the order
\begin{equation}
\mathcal{B}_{eq}\sim \mathcal{H}_{M,0}^{1/2}\Delta_{r-e}^{-1/3}\sim 
\frac{1}{\pi^{2/3}e^{5/3}} \left(\frac{H}{m}\right)^{1/2}
\left(\frac{h_{eq}}{h_{rh}}\right)^{-1/6}\label{e-8}
\end{equation}
with a corresponding (dimensionless) comoving coherence length $\ell_{eq}\sim \Delta_{r-e}^{2/3} = \left(h_{eq}/h_{rh}\right)^{1/3}$.
To obtain the value of the physical field and coherence scale we must take into account the expansion of the Universe, which for the
field means a dilution by a factor of $\left( h_{rh}/h_{eq}\right)$ and for the coherence scale an expansion by a factor 
$\left( h_{eq}/h_{rh}\right)^{1/2}$. Adding also the dimensions through the corresponding powers of $H$ we finally have
$B_{eq} \sim H^2\times 10^{20}\left(\mathrm{Gauss}/\mathrm{GeV}^2\right)\left( h_{rh}/h_{eq}\right)\mathcal{B}_{eq}$
and $\lambda_{eq} = H^{-1}\times 6.4\times 10^{-39}\left(\mathrm{Mpc}.\mathrm{GeV}\right)\left( h_{eq}/h_{rh}\right)^{1/2}\ell_{eq}$. 
Using the figures given above and in Subsection \ref{dim} we obtain at the beginning of matter dominance, a field intensity in the range
\begin{equation}
10^{-9}\mathrm{Gauss} \lesssim B_{eq}\lesssim 10^{-5}\mathrm{Gauss} \label{e-9}
\end{equation}
with a corresponding coherence length of
\begin{equation}
10^{-3}\mathrm{pc} \lesssim \lambda_{eq} \lesssim 10^{-1} \mathrm{pc} \label{e-10}
\end{equation}
These values could impact the process of early structure formation \cite{ryu-12}.

\section{\label{dc}Discussion and Conclusions}

In this work we have investigated the generation of magnetic helicity in primordial magnetogenesis. 
We considered a specific mechanism for magnetic field generation, developed in Refs. \cite{ckm-98,giov-shap-00} 
(see also \cite{cal-hu-08}), where stochastic magnetic fields were induced by electric currents that appeared
due to particle creation at the transition Inflation-Reheating. The charges correspond to a scalar field
minimally coupled to gravity, as in this case the number of particles created is maximal. There would be support
for such a field in the Supersymmetric theory of particles \cite{kcmw-00}.
As the induced magnetic fields are random, the mean values of the different quantities are null, and consequently
we had to calculate a rms deviation of $H_M$. We could write it as the sum of four different SQED Feynman graphs
of differente multiplicities. 

Our main result is that the fields induced by stochastic currents generated by cosmological particle creation are helical.

On large scales as e.g. the galactic ones, we considered that the evolution of the magnetic helicity is diffusive because
that scales are larger than the horizon during magnetogenesis, entering the horizon by the end of the radiation era or
during matter dominance. 

We also investigated the generation of magnetic helicity at a scales with a size of the order of the horizon during reheating, 
where inverse cascade can be operative. We made the naive hypothesis that throughout radiation dominance the plasma flow
is endowed with decaying turbulence and applied the model for magnetic energy and coherence scale evolution developed in Refs.
\cite{biskamp-03,kahniashvili-13}. The estimation of the resulting magnetic field intensities at equilibrium between matter
and radiation gives values and coherence scales that could be of importance for structure formation \cite{ryu-12}.

The importance of primordial magnetic fields endowed with magnetic helicity is that fields coherent on scales equal or
shorter than the particle horizon, would self-organize on larger scales. This is due to the fact that when the plasma where the field evolves possesses some degree of (decaying) turbulence, magnetic helicity performs an inverse cascade instead of a direct one, 
thus thus self-organizing at large scales. For large scales, as e.g. the galactic ones, there can also operate the inverse cascade, 
but even if there were not such a process, the operation of further amplifying mechanisms, as galactic dynamos, would be crucially 
affected by the topological properties of the seed fields \cite{blackman}.

In conclusion, generation of magnetic helicity in the early Universe seems to be quite easily achieved in different scenarios  
\cite{corn-97,son-99,vachas-01,copi-08,chu-11,long-13}. The problem still remains in the intensities. In the mechanism considered 
here the fields obtained are indeed helical, but their intensities on large scales are too small, or have a marginal value to have an 
astrophysical impact, while on smaller scales they can be important provided that inverse cascade of magnetic helicity is operative
in the early universe.

\section{Acknowledgements}
E.C. aknowledges suport from CONICET, UBA and ANPCyT. A. K. thanks CNPq/CAPES for financial help through the 
PROCAD project 552236/2011-0.

\appendix

\section{\label{apa}Solutions of Klein-Gordon Equation and Calculation of the Bogoliubov Coefficients}

In this appendix we find the solutions of eq. (\ref{a-5b}) in the two
epochs of the universe considered in the paper, i.e., Inflation and Reheating,
and calculate the Bogoliubov coefficients. The Fourier transformed eq. (\ref{a-5b}) reads
\begin{equation}
\partial _{\tau }^{2}\varphi _{p}+\left[ p^{2}+a^{2}\left( \tau \right) 
\frac{m^{2}}{H^{2}}-\frac{\ddot{a}\left( \tau \right) }{a\left( \tau \right) 
}\right] \varphi _{p}=0  \label{apa-1}
\end{equation}
whose solutions in each epoch will be labeled as $\varphi_p^I\left(\tau\right)$ (Inflation) and
$\varphi_p^R\left(\tau\right)$ (Reheating). The fact that they are different in each epoch tells
that a given quantum state in the first epoch will not coincide with a same state in the subsequent 
epoch. More concretely, a vacuum state during Inflation will appear as a particle state in Reheating. 
Mathematically this is expressed as
\cite{birrel-94}
\begin{equation}
\varphi_p^I \left(\tau\right) = \alpha_p\varphi_p^R \left(\tau\right) + \beta_p\varphi_p^{\ast R}
\left(\tau\right)\label{apa-0a}
\end{equation}
The fact that $\beta_p \not= 0$, shows that the two quantum states are not equivalent. $\alpha_p$ 
and $\beta_p$ are the Bogoliubov coefficients \cite{birrel-94}. For modes corresponding to scales 
larger than the horizon size at the epoch of transition, that transition can be considered as 
instantaneous and the coefficients can be calculated by demanding continuity of $\varphi_p^I$ and 
$\dot \varphi_p^I$ at that instant. This gives
\begin{eqnarray}
\alpha_p &=& \frac{\varphi_p^I\left(0\right)\dot\varphi_p^{\ast R}\left(0\right)
-\varphi_p^{\ast R}\left(0\right)\dot\varphi_p^I\left(0\right)}
{\varphi_p^{R}\left(0\right)\dot\varphi_p^{\ast R}\left(0\right)-\varphi_p^{\ast R}\left(0\right)
\dot\varphi_p^{R}}\left(0\right)\label{apa-0b}\\
\beta_p &=& \frac{\varphi_p^R\left(0\right)\dot\varphi_p^{I}\left(0\right)-\varphi_p^{I}\left(0\right)
\dot\varphi_p^R\left(0\right)}
{\varphi_p^{R}\left(0\right)\dot\varphi_p^{\ast R}\left(0\right)-\varphi_p^{\ast R}\left(0\right)
\dot\varphi_p^{R}\left(0\right)}\label{apa-0c}
\end{eqnarray}
For scales shorter than the horizon, i.e., $p> 1$, the calculation is more involved, 
as details of the transition do matter (see Ref. \cite{cal-kan-10} and references therein). For completion 
we only quote here the result and refer the reader to the references for the calculations. 
\begin{equation}
\beta_p \simeq \frac{i}{16\tau_0}\frac{\exp\left[i\tau_0S\left(0\right)\right]}{p^5},\qquad p > 1 \label{apa-0d}
\end{equation}
where $\tau_0$ is the lasting of the transition.

\subsection{Solutions during Inflation}

Considering that the scale factor at the end of Inflation is equal to unity, we can
write $a\left( \tau \right) =\left( 1-\tau \right) ^{-1}$ and eq. 
(\ref{apa-1}) reads
\begin{equation}
\left[ \frac{\partial ^{2}}{\partial \tau ^{2}}+p^{2}+\frac{m^{2}}{H^{2}}
\left( 1-\tau \right) ^{2}-\frac{2}{\left( 1-\tau \right) ^{2}}\right]
\varphi _{p}^{I}=0  \label{apa-2}
\end{equation}
Writing $\varphi _{p}^{I}=\left( 1-\tau \right) ^{1/2}f_{p}$ eq. (\ref{apa-2}
) transforms into a Bessel equation for $f_{p}$ and therefore the
normalized, positive frequency solutions of eq. (\ref{apa-2}) are
\begin{equation}
\varphi_{p}^{I}=\frac{\pi ^{1/2}}{2}\left( 1-\tau \right) ^{1/2}H_{\nu
}^{\left( 1\right) }\left[ p\left( 1-\tau \right) \right]  \label{apa-3}
\end{equation}
with $\nu =\sqrt{9/4-m^{2}/H^{2}}$. As we shall be considering the case $m/H \ll 1$
we can approximate $\nu \simeq 3/2$ and in this case the Hankel function has a
polinomic expression, namely
\begin{equation}
\varphi_{p}^{I}\simeq -\frac{e^{ik\left( 1 - \tau\right)}}{\sqrt{2k}}
\left[1+\frac{i}{k\left( 1 - \tau\right)}\right] \label{apa-4}
\end{equation}

\subsection{Solutions during Reheating}

In this case $a\left( \tau \right) =\left( 1+\tau /2\right) ^{2}$ and eq. (
\ref{apa-1}) is given by
\begin{equation}
\left[ \frac{\partial ^{2}}{\partial \tau ^{2}}+p^{2}+\frac{m^{2}}{H^{2}}
\left( 1+\frac{\tau }{2}\right) ^{4}-\frac{1/2}{\left( 1+\tau /2\right) ^{2}}
\right] \varphi _{p}^{R}=0  \label{apa-5}
\end{equation}
which cannot be reduced to a known equation, unless some approximations are
made. In this sense we consider two situations: $p < m/H$ and $m/H < p < 1$.
In both cases the modes correspond to wavelengths larger than the size of 
the particle horizon and hence detais of the transition between the two
considered epochs do not matter. The corresponding Bogoliubov coefficients
can be calculated considering an instantaneous transition at $\tau = 0$,
and demanding continuity of the modes and their first derivatives at that moment.

\subsubsection{Limit $p\leq m/H$.}

In this limit eq. (\ref{apa-5}) reads
\begin{equation}
\left[ \frac{\partial ^{2}}{\partial \tau ^{2}}+\frac{m^{2}}{H^{2}}\left( 1+
\frac{\tau }{2}\right) ^{4}-\frac{1/2}{\left( 1+\tau /2\right) ^{2}}\right]
\varphi _{p}^{R}=0  \label{apa-6}
\end{equation}
Proposing a solution of the form $\varphi _{p}^{R}\left(\tau\right) =\left( 1+\tau /2\right)
^{1/2}f_{p}\left[ \left( 1+\tau /2\right) ^{3}\right] $ eq. (\ref{apa-7})
transforms into a Bessel equation for $f_{p}$, and so the normalized,
positive frequency solutions of eq. (\ref{apa-6}) are
\begin{eqnarray}
\varphi^{R}\left( p\right)&=&i\sqrt{\frac{H}{2m}}\left( 1+\frac{\tau }{2}\right)
^{1/2}H_{1/2}^{\left( 2\right) }\left[ \frac{2m}{3H}\left( 1+\frac{\tau }{2}
\right) ^{3}\right] \nonumber\\
&=& -\sqrt{\frac{6}{\pi}}\frac{\exp\left[i\left( 2m/3H\right)\left(1 + 
\tau/2\right)^3\right]}{\left( 1+\tau/2\right)}\label{apa-7b}
\end{eqnarray}
Using expr. (\ref{apa-0c}) we obtain
\begin{equation}
\beta_p \simeq -i\sqrt{\frac{H}{2m}}\sqrt{\frac{9}{8}}p^{-3/2}\label{apa-8}
\end{equation}

\subsubsection{Limit $m/H < p < 1$.}

This case still corresponds to modes outside the particle horizon, but their
form is different from the one corresponding to the previous momentum interval. 
Eq. (\ref{apa-4}) reads
\begin{equation}
\left[ \frac{\partial ^{2}}{\partial \tau ^{2}}+p^{2}-\frac{1/2}{\left( 1+\tau /2\right) ^{2}}
\right] \varphi _{p}^{R}=0  \label{apa-9}
\end{equation}
Writing $\varphi_p^{R}\left(\tau\right) = \left( 1 + \tau/2\right)^{1/2}g_p
\left[2p\left(1 + \tau/2\right)\right]$ we again obtain a Bessel equation for $g_p$.
In this case the normalized, positive frequency modes are
\begin{eqnarray}
\varphi_p^{R}\left(\tau\right)&=&\sqrt{\frac{\pi}{2}}\left(1+\frac{\tau}{2}\right)^{1/2}
H_{3/2}^{(2)}\left[2p\left(1+\frac{\tau}{2}\right)\right]\nonumber\\
&=& -\frac{1}{\sqrt{2p}}e^{i2p\left(1+\tau /2\right)}\left[1-\frac{i}{4p\left(1+\tau /2\right)}\right]
\label{apa-10}
\end{eqnarray}
Replacing again in expr. ({apa-0c}) we have
\begin{equation}
\beta_p \simeq -i\frac{3}{8}e^{-ip}p^{-3}\label{apa-11}
\end{equation}

\section{\label{apb}Retarded propagator for the electromagnetic field}

The Fourier transform of the retarded propagator for the electromagnetic
field satisfies the following equation

\begin{equation}
\left[ \partial _{\tau }^{2}+\frac{\sigma _{0}}{H}\partial _{\tau }+\kappa
^{2}\right] G_{ret}\left( \kappa ,\eta ,\tau \right) =\delta \left( \eta
-\tau \right)  \label{apb-1}
\end{equation}%
whose homogeneous solutions are of the form $\exp \left( -\sigma _{0}\tau
/2H\right) \exp \left( \pm \alpha \tau \right) $, with $\alpha =\sqrt{\sigma
_{0}^{2}/4H^{2}-\kappa ^{2}}$. We therefore propose 
\begin{equation}
G_{ret}\left( \kappa ,\eta ,\tau \right) =Ae^{-\sigma _{0}\left( \eta -\tau
\right) /2H}\left[ e^{\alpha \left( \eta -\tau \right) }-e^{-\alpha \left(
\eta -\tau \right) }\right] \Theta \left( \eta -\tau \right)  \label{apb-2}
\end{equation}%
which is continuous in $\eta \rightarrow \eta ^{\prime }$. Demanding $\left.
dG_{ret}\left( \kappa,\eta ,\eta ^{\prime }\right) /d\eta \right\vert _{\eta
=\eta ^{\prime }}=-1$ gives $A=1/2\alpha $, whence
\begin{equation}
G_{ret}\left( \kappa ,\eta ,\tau \right) =e^{-\sigma _{0}\left( \eta -\tau
\right) /2H}\left[ \frac{e^{\alpha \left( \eta -\tau \right) }-e^{-\alpha
\left( \eta -\tau \right) }}{2\alpha }\right] \Theta \left( \eta -\tau
\right)  \label{apb-3}
\end{equation}

\subsection{Limiting form for $\kappa < \sigma_0/2H$}
In this case we can take $\alpha \simeq \sigma_0/2H$ and thus
\begin{equation}
G_{ret}\left(\kappa ,\eta ,\tau \right) \simeq \frac{H}{\sigma_0}\left[1-
e^{-\sigma_0\left(\eta - \tau\right)/H}\right]\Theta \left( \eta -\tau
\right) \label{apb-4}
\end{equation}
In the limit $\eta \rightarrow \infty$ eq. (\ref{apb-4}) simply reads
\begin{equation}
G_{ret}\left(\kappa ,\eta ,\tau \right)\simeq \frac{H}{\sigma_0} \Theta \left( \eta -\tau
\right)\label{apb-5}
\end{equation}
while for $\eta - \tau \rightarrow 0$ we have
\begin{equation}
G_{ret}\left(\kappa ,\eta ,\tau \right)\simeq \left(\eta - \tau\right)\Theta\left(\eta - \tau\right)
\label{apb-6}
\end{equation}

\subsection{Limiting form for $\kappa > \sigma_0/2H$}
In this case $\alpha \simeq i\kappa$ and then

\begin{equation}
G_{ret}\left( \kappa,\eta ,\tau\right)\simeq e^{-\sigma_0\left(\eta - \tau\right)/2H}
\frac{\sin\left[\kappa\left(\eta - \tau\right)\right]}{\kappa}
\Theta \left( \eta -\tau \right) \label{apb-7}
\end{equation}
which for finite $\kappa$ and $\left(\eta - \tau\right) \rightarrow 0$ gives
\begin{equation}
G_{ret}\left( \kappa,\eta ,\tau\right)\simeq e^{-\sigma_0\left(\eta - \tau\right)/2H}
\left( \eta -\tau \right) \Theta \left( \eta -\tau \right) \label{apb-8}
\end{equation}

\section{\label{apc}Contribution from the "cross" and "two bubbles" diagrams}

We must evaluate
\begin{equation}
A^{-1}B\left[ C-B\right]  \label{apd-1}
\end{equation}%
with%
\begin{equation}
A=q_{1}^{2\nu }q_{3}^{2\nu }\left\vert \bar{p}-\bar{q}_{1}\right\vert ^{2\nu
}\left\vert \bar{\kappa}-\bar{p}-\bar{q}_{3}\right\vert ^{2\nu }
\label{apd-2}
\end{equation}%
\begin{equation}
B=\left( 2\bar{q}_{1}-\bar{p}\right) \cdot \left( \bar{\kappa}\times \bar{q}%
_{3}\right) -2\bar{q}_{1}\cdot \left( \bar{p}\times \bar{q}_{3}\right)
\label{apd-3}
\end{equation}%
\begin{equation}
C=\left( \bar{p}-\bar{q}_{3}\right) \cdot \left( \bar{\kappa}\times \bar{q}%
_{1}\right) -2\bar{q}_{1}\cdot \left( \bar{p}\times \bar{q}_{3}\right)
\label{apd-4}
\end{equation}%
We have that%
\begin{equation}
C-B=\bar{p}\cdot \left( \bar{\kappa}\times \bar{q}_{1}\right) +\bar{q}%
_{3}\cdot \left( \bar{\kappa}\times \bar{q}_{1}\right) +\bar{p}\cdot \left( 
\bar{\kappa}\times \bar{q}_{3}\right)  \label{apd-5}
\end{equation}%
and 
\begin{equation}
B=2\bar{\kappa}\cdot \left( \bar{q}_{3}\times \bar{q}_{1}\right) +\bar{\kappa%
}\cdot \left( \bar{p}\times \bar{q}_{3}\right) +2\bar{p}\cdot \left( \bar{q}%
_{1}\times \bar{q}_{3}\right)  \label{apd-6}
\end{equation}%
Defining $\bar{p}=\bar{q}+\bar{q}_{1}$ we can write%
\begin{equation}
C-B=\bar{q}\cdot \left( \bar{\kappa}\times \bar{q}_{1}\right) +\bar{q}\cdot
\left( \bar{\kappa}\times \bar{q}_{3}\right)  \label{apd-7}
\end{equation}%
and%
\begin{eqnarray}
B &=&\bar{\kappa}\cdot \left( \bar{q}_{3}\times \bar{q}_{1}\right) +\bar{%
\kappa}\cdot \left( \bar{q}\times \bar{q}_{3}\right) +2\bar{q}\cdot \left( 
\bar{q}_{1}\times \bar{q}_{3}\right)  \label{apd-8} \\
&\equiv &B_{1}+B_{2}+B_{3}  \notag
\end{eqnarray}%
For the different terms in the integrand we thus have%
\begin{equation}
B_{1}\left( C-B\right) =\left[ \bar{\kappa}\cdot \left( \bar{q}_{3}\times 
\bar{q}_{1}\right) \right] \left[ \bar{\kappa}\cdot \left( \bar{q}_{1}\times 
\bar{q}\right) \right] +\left[ \bar{\kappa}\cdot \left( \bar{q}_{3}\times 
\bar{q}_{1}\right) \right] \left[ \bar{\kappa}\cdot \left( \bar{q}_{3}\times 
\bar{q}\right) \right]  \label{apd-9}
\end{equation}%
that integrates to zero because the first term is odd in $\bar{q}_{3}$ and
the second in $\bar{q}_{1}$. For the same reasons, the same happens with the
term%
\begin{equation}
B_{3}\left( C-B\right) =2\left[ \bar{q}\cdot \left( \bar{q}_{1}\times \bar{q}%
_{3}\right) \right] \left[ \bar{\kappa}\cdot \left( \bar{q}_{1}\times \bar{q}%
\right) \right] +2\left[ \bar{q}\cdot \left( \bar{q}_{1}\times \bar{q}%
_{3}\right) \right] \left[ \bar{\kappa}\cdot \left( \bar{q}_{3}\times \bar{q}%
\right) \right]  \label{apd-10}
\end{equation}%
There remains the term%
\begin{equation}
B_{2}\left( C-B\right) =\left[ \bar{\kappa}\cdot \left( \bar{q}\times \bar{q}%
_{3}\right) \right] \left[ \bar{\kappa}\cdot \left( \bar{q}_{1}\times \bar{q}%
\right) \right] -\left[ \bar{\kappa}\cdot \left( \bar{q}\times \bar{q}%
_{3}\right) \right] ^{2}  \label{apd-11}
\end{equation}
We define
\begin{equation}
\bar{Q}_{+}=\frac{\bar{q}_{1}+\bar{q}_{3}}{2},\qquad \bar{Q}_{-}=\frac{\bar{q%
}_{1}-\bar{q}_{3}}{2}  \label{apd-12}
\end{equation}
hence
\begin{equation}
B_{2}\left( C-B\right) =-2\left[ \bar{\kappa}\cdot \left( \bar{q}\times \bar{%
Q}_{+}\right) \right] ^{2}+2\left[ \bar{\kappa}\cdot \left( \bar{q}\times 
\bar{Q}_{+}\right) \right] \left[ \bar{\kappa}\cdot \left( \bar{q}\times 
\bar{Q}_{-}\right) \right]  \label{apd-13}
\end{equation}%
The term odd in $\bar{Q}_{\pm }$ integrate to zero, hence there only remains
the first one, i.e.%
\begin{equation}
B_{2}\left( C-B\right) =-2\left[ \bar{\kappa}\cdot \left( \bar{q}\times \bar{%
Q}_{+}\right) \right] ^{2}  \label{apd-14}
\end{equation}%
which is clearly non-null.

\end{document}